\documentclass{article}

\usepackage{tikz, pgf, pgfplots}
\usetikzlibrary
	{
		arrows, automata, chains, datavisualization.formats.functions, fit, positioning, shapes
	}

\usepackage{amsthm}
\usepackage{amssymb}
\usepackage[bb = boondox]{mathalfa}
\usepackage{dsfont}
\usepackage{mathtools}
\usepackage{multicol}
\usepackage{braket}
\usepackage{physics}
\numberwithin{equation}{section}

\usepackage{yquant, braket}
\usetikzlibrary{quantikz, fit}

\usepackage[a4paper, left = 2.5cm, right = 2.5 cm, top = 2.5cm, bottom = 3.0 cm]{geometry}

\usepackage{xcolor}
\usepackage{varwidth}
\usepackage[breakable, skins, theorems]{tcolorbox}
\usepackage{graphicx}
\usepackage{hyperref}
\hypersetup
{
	colorlinks,
	linkcolor = {WordRed!75!black},
	citecolor = {WordBlueDarker25},
	urlcolor = {WordAquaDarker50}
}
\usepackage{nicematrix}
\NiceMatrixOptions
{
	code-for-first-row = \color{RedPurple},
	code-for-last-col = \color{WordAquaDarker25}
}
\usepackage{colortbl}
\usepackage{float}
\usepackage{cellspace}
\usepackage{makecell}
\usepackage[boxed, ruled, vlined]{algorithm2e}
\usepackage{appendix}
\usepackage{multirow}
\newcommand{\orcidicon}[1]{\href{https://orcid.org/#1}{\includegraphics[height=\fontcharht\font`\B]{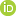}}}

\newtheorem{example}{Example}[section]

\definecolor{MyLightRed}{RGB}{244, 213, 245}
\definecolor{WordRed}{RGB}{255, 0, 102}
\definecolor{RedDarkLightest}{HTML}{ff0088}
\definecolor{RedDarkLight}{HTML}{ea005f}
\definecolor{RedPurple}{HTML}{aa007f}
\definecolor{Purple}{HTML}{911146}
\definecolor{WordLightGreen}{RGB}{140, 214, 192}
\definecolor{WordGreen}{RGB}{0, 176, 80}
\definecolor{GreenLightest}{HTML}{00ffa0}
\definecolor{GreenLighter1}{HTML}{00b383}
\definecolor{GreenLighter2}{HTML}{00aa7f}
\definecolor{GreenDark}{HTML}{225522}
\definecolor{GreenTeal}{HTML}{008080}
\definecolor{WordIceBlue}{RGB}{223, 227, 229}
\definecolor{MyVeryLightBlue}{RGB}{211, 245, 247}
\definecolor{WordBlueVeryLight}{RGB}{0, 176, 240}
\definecolor{WordBlueLight}{RGB}{0, 112, 192}
\definecolor{WordBlueDark}{RGB}{46, 116, 181}
\definecolor{WordBlueDarker}{RGB}{31, 78, 121}
\definecolor{WordBlueDarker25}{RGB}{54, 96, 146}
\definecolor{WordBlueDarker50}{RGB}{36, 64, 98}
\definecolor{WordBlueDarkest}{RGB}{0, 32, 96}
\definecolor{WordBlue}{RGB}{19, 65, 99}
\definecolor{MyBlue}{RGB}{0, 64, 128}
\definecolor{MyDarkBlue}{RGB}{0, 51, 102}
\definecolor{BlueVeryDark}{HTML}{222255}
\definecolor{WordAquaLighter80}{RGB}{218, 238, 243}
\definecolor{WordAquaLighter60}{RGB}{183, 222, 232}
\definecolor{WordAquaLighter40}{RGB}{146, 205, 220}
\definecolor{WordAquaDarker25}{RGB}{49, 134, 155}
\definecolor{WordAquaDarker50}{RGB}{33, 89, 103}
\definecolor{WordVeryLightTeal}{RGB}{223, 236, 235}
\definecolor{WordLightTeal}{RGB}{160, 199, 197}
\definecolor{WordDarkTealLighter80}{RGB}{207, 223, 234}
\definecolor{WordDarkTeal}{RGB}{72, 123, 119}
\definecolor{WordDarkerTeal}{RGB}{48, 82, 80}
\definecolor{WordTurquoiseLighter80}{RGB}{209, 238, 249}

\title
	{
		Experimental analysis of quantum annealers and hybrid solvers using benchmark optimization problems
	}

\author
	{
		Evangelos Stogiannos$^1$\orcidicon{0000-0001-7661-9342}
		and
		Christos Papalitsas$^1$\orcidicon{0000-0003-0467-796X}
		and
		Theodore Andronikos$^1$\orcidicon{0000-0002-3741-1271} \\
		$^1$Department of Informatics, Ionian University, \\
		7 Tsirigoti Square, 49100 Corfu, Greece; \\
		\{p18stog, papalitsas, andronikos\}@ionio.gr \\
	}

\begin{document}

\maketitle

\begin{abstract}
	This paper studies the Hamiltonian Cycle Problem (HCP) and the Traveling Salesman Problem (TSP) on D-Wave's quantum systems. Initially, motivated by the fact that most libraries present their benchmark instances in terms of adjacency matrices, we develop a novel matrix formulation for the HCP and TSP Hamiltonians, which enables the seamless and automatic integration of benchmark instances in quantum platforms. our extensive experimental tests have led us to some interesting conclusions. D-Wave's {\tt Advantage\_system4.1} is more efficient than {\tt Advantage\_system1.1} both in terms of qubit utilization and quality of solutions. Finally, we experimentally establish that D-Wave's Hybrid solvers always provide a valid solution to a problem, without violating the QUBO constraints, even for arbitrarily big problems, of the order of $120$ nodes. When solving TSP instances, the solutions produced by the quantum annealer are often invalid, in the sense that they violate the topology of the graph. To address this use we advocate the use of \emph{min-max normalization} for the coefficients of the TSP Hamiltonian. Finally, we present a thorough mathematical analysis on the precise number of constraints required to express the HCP and TSP Hamiltonians. This analysis, explains quantitatively why, almost always, running incomplete graph instances requires more qubits than complete instances. It turns out that incomplete graph require more quadratic constraints than complete graphs, a fact that has been corroborated by a series of experiments.

\end{abstract}
\section{Introduction} \label{sec:Introduction}

Quantum computers are the strongest contenders against conventional, silicon-based systems and gain traction by the day. Proposed by Richard Feynman in 1982, the idea of using quantum ``parallelism'' \cite{Feynman1982} to solve computational problems has been proven true in quite a few cases, like Shor's \cite{Shor1999} and Grover's \cite{Grover1996} algorithms, which are more efficient than their conventional counterparts. Further information about quantum computing and information is available in \cite{Nielsen2010}.

The adiabatic theorem is a staple in quantum mechanics \cite{Farhi2000,Farhi2001}. According to the theorem, when a quantum system with a starting ground state of $H_{init}$ is subject to a gradually changing Hamiltonian, with the same starting ground state and a final ground state of $H_{fin}$, the system itself will adapt to the changes, ending at $H_{fin}$. In the first decade of the 2000s, Farhi et al. introduced adiabatic quantum computation \cite{Messiah1961,Amin2009}, basing it on the aforementioned adiabatic theorem. Adiabatic quantum computation has been shown to be equivalent to standard quantum computation \cite{Aharonov2008}.

In the early 2000s, Kadowaki and Nishimori introduced quantum annealing, a metaheuristic based on the principles of quantum adiabatic computation \cite{Kadowaki1998}. Since then, it has gained prominence for the tackling of combinatorial optimization problems. Quantum annealing, similar to conventional simulated annealing \cite{Kirkpatrick1983}, uses a multivariable function to create an energy landscape of all the possible states in a quantum superposition, with its ground state representing the problem's solution. Then, the system goes through the quantum annealing process repeatedly until it encounters an optimal solution with a high probability. The difference between quantum and simulated annealing is the multi-qubit tunneling used by quantum annealing \cite{Kadowaki1998}, which also improves performance due to the high degree of parallelism. Quantum annealers try to calculate the optimal solutions by analyzing every possible input simultaneously, an attribute that might be vital when handling NP-complete problems. However, as of now, these methods are still in development. Thus, it is recommended for them to be used as an automatic heuristic-finding program than as a precise solving program \cite{Pakin2018}.

When solving large scale optimization problems, one of the most effective classical strategies is to search among the nearest neighbors and search for paths between the initial and final configurations to improve upon an initial guess $y_t$. This search takes place locally among neighboring configurations similar to $y_t$. By finding the best solution within the local neighborhood of $y_{t}$, the next candidate solution $y_{t+1}$ appears. Then a new local search starts with $y_{t+1}$ as the starting point in the neighborhood. Unfortunately, greedy local improvement, in case of hard problems, may be deceptively leading the solution into a local minima whose energy may be much higher than the globally minimum value.

Optimization problems are one of the areas where using quantum computers is considered to be advantageous \cite{PerdomoOrtiz2012, Biamonte2017, Papalitsas2018, Rebentrost2019}. To solve an optimization problem with a quantum computer, we first need to create a Hamiltonian, the ground state of which represents an optimal solution for the problem. In most cases using quantum annealing, a system starts in an equal superposition for all its states with said Hamiltonian applied. Over time the system evolves according to the  Schr\"{o}dinger equation, and the system's state changes depending on the strength of the local transverse field as it changes over time. To get the problem's solution, we slowly turn off the transverse field and, the system settles in its ground state. If we have chosen the correct Hamiltonian, then the ground state of our system will also be its optimal solution.

D-Wave offers quantum computing systems using quantum annealing for solving optimization and probabilistic sampling problems. One of the optimization problems that the D-Wave computers have demonstrated the ability to solve is the quadratic unconstrained binary optimization problem. The quantum processing units (QPUs) inside the D-Wave machines handle the quantum annealing process. The lowest energy states of the superconducting loops are the quantum bits (which we shall refer to as qubits) and are the analog of conventional bits for the QPU \cite{Choi2008, Boothby2019}.
The quantum annealing process takes a system with all of its qubits in superposition and guides it towards a new state in which every qubit collapses into a value of either $0$ or $1$. The resulting system will be a system in a classical state, which will also be the optimal solution to the problem. In 2020 D-Wave announced their new generation of quantum computers, the Advantage System. The Advantage System uses the Pegasus topology, improving from the previous generation's Chimera topology. The Advantage QPU contains more than $5000$ qubits, with each qubit having $15$ couplers to other qubits, totaling more than $35000$ couplers. Qubits in the Advantage QPU are mapped to a P16 Pegasus graph, meaning they are logically mapped into a $15 \times 15 \times 2$ matrix of unit cells on a diagonal grid. A Pegasus unit cell contains twenty-four qubits, with each qubit coupled to one similarly aligned qubit in the cell and two similarly aligned qubits in adjacent cells \cite{D-WaveQPUTopology2022}. The percentage of total working qubits is called the working graph and is a subgraph of the total connected cells that physically exist in the QPU.

One of the types of problems that D-Wave computers can solve is the \emph{quadratic unconstrained binary optimization} (QUBO for short) problem. A QUBO model is a pattern matching technique with many applications, ranging from machine learning to solving optimization problems \cite{UshijimaMwesigwa2017}. Its basis is minimizing a quadratic polynomial function over binary variables \cite{Boros1990, Lewis2017, Kochenberger2014, Lloyd2013, Hauke2019, Boothby2019, Glover2018, Boros1990} and thus is an NP-hard problem \cite{Boros2007}. The Schr\"{o}dinger equation is a linear partial differential equation. It expresses the wave function or state function of a quantum-mechanical system. By knowing the necessary variables in a system, we can model most physical systems. The Ising model is a model that is equivalent to the QUBO model. Proposed in the 1920s by Ernst Ising and Wilhelm Lenz, the Ising model is a well-researched model in ferromagnetism \cite{Newell1953, Lucas2014}. It assumes that qubits represent the model's variables, and their interactions stand for the costs associated with each pair of qubits. The model's design makes it possible to formulate it into an undirected graph with qubits as vertices and couplers as edges among them. In 2017, D-Wave introduced qbsolv, an open-source software that, according to D-Wave, can handle problems of arbitrary size. Qbsolv is a hybrid system meaning it breaks down the problem's graph into smaller subgraphs using its CPU and solves each partition using its QPU. It then repeats the process and uses a Tabu-based search to examine if a better solution is available \cite{Neukart2017}.

\subsection{Related work} \label{subsec:Related Work}

The Hamiltonian Cycle Problem (HCP for short) asks a simple yet complex in answering question: ``Given a graph of n nodes, is there a path that passes through each node only once and returns to the starting node?'' It is an NP-Complete problem, first proposed in the 1850s, and is a fascinating problem for computer scientists. Flinders Hamiltonian Cycle Project (FHCP) provides many benchmarking graph datasets for both the HCP and TSP \cite{FHCP2022}. In this paper we have used their dataset of small graphs, generated by GENREG \cite{Meringer1999}. Their small number of nodes made them suitable for solving by the QPU without needing a hybrid solver. The Traveling Salesman Problem (TSP from now on) is an NP-hard combinatorial optimization problem that builds upon the HCP \cite{Shmoys1985}. Given a weighted graph, instead of finding the existence of a Hamiltonian Cycle, the TSP asks what's the cycle with the minimum total cost.

The complexity of the TSP makes it a great topic of interest for many researchers, and leads them to pursue other avenues. One such alternative approach advocates the use of metaheuristics, i.e., high-level heuristics designed to select a lower-level heuristic that can produce a fairly good solution with limited computing capacity \cite{bianchi_survey_2009}. The term ``metaheuristics'' was suggested by Glover. Of course metaheuristic procedures, in contrast to exact methods, do not guarantee a global optimal solution \cite{blum2003metaheuristics}. Papalitsas et al. \cite{Papalitsas2015} designed a metaheuristic based on VNS for the TSP with emphasis on Time Windows. This quantum-inspired procedure was also applied successfully to the solution of real-life problems that can be modeled as TSP instances \cite{Papalitsas2018}. More recently, \cite{Papalitsas2019b} applied a quantum-inspired metaheuristic for tackling the practical problem of garbage collection with time windows that produced particularly promising experimental results, as further comparative analysis demonstrated in \cite{Papalitsas2019c}. A thorough statistical and computational analysis on asymmetric, symmetric, and national TSP benchmarks from the well known TSPLIB benchmark library was conducted in \cite{Papalitsas2019a}.

Another state-of-the-art approach for combinatorial optimization problem like TSP is to employ unconventional computing \cite{Gan2010, Aono2011}. It turns out that HCP and TSP are suitable candidates for formulation into the QUBO model so as to be solved by a quantum annealer. An excellent reference for many Ising formulations for NP-complete and NP-hard optimization problems is the paper by Lucas \cite{Lucas2014}. In this paper, Lucas also presented the QUBO formulations for the Hamiltonian Cycle Problem and the Traveling Salesman Problem, giving particular emphasis to the minimization of the usage of qubits. Another formulation of the Traveling Salesman Problem, focusing on handling the symmetric version of the problem, was presented by Martonák et al. \cite{Martonak2004}. The basis for the QUBO formulation used in this paper was proposed by Papalitsas et al. in \cite{Papalitsas2019}. A recent work \cite{Warren2020} conducted an experimental analysis of the performance of two hybrid systems provided by D-Wave, the Kerberos solver and the LeapHybridSampler solver, using the TSP as a benchmark. The conclusion of the tests in \cite{Warren2020} is that the Kerberos solver is superior to the LHS, as Kerberos consistently yields solutions closer to the optimal route for every TSP instance. However, the LHS still produced routes that obeyed the imposed constraints for the TSP, even if the quality of its solutions wasn't as high as Kerberos's.

\textbf{Contribution}. Motivated by the fact that most libraries present their benchmark instances in terms of adjacency matrices, and in order to facilitate their execution by the quantum annealer and by tools like \texttt{qubovert} \cite{Qubovert2022}, we set out convert the HCP and TSP Hamiltonians in matrix form. This formulation, which we believe has not appeared previously in the literature, enables the seamless and automatic integration of the benchmark instances in quantum testing platforms.

We also present a thorough mathematical analysis on the precise number of constraints required to express the HCP and TSP Hamiltonians. This analysis, explains quantitatively why, almost always, running incomplete graph instances requires more qubits than complete instances. It turns out that incomplete graph require more quadratic constraints than complete graphs. This theoretical finding has been corroborated by a series of experiments outlined in subsection \ref{subsec:The Effect of Graph Connectivity on Qubit Usage}. Its importance is not only theoretical, but practical as well, since in the current technological stage, qubits still remain a precious commodity.

When solving TSP instances, the solutions produced by the quantum annealer are often invalid, in the sense that they violate the topology of the graph. This is to be expected because if the weight of an edge is higher than the constraint penalty, the annealer, in order to reach a lower energy state, ignores the constraint. To address this use we advocate the use of \emph{min-max normalization} of the coefficients of the TSP Hamiltonian. This well-known and easy to implement technique was experimentally tested and found to be considerably helpful.

Furthermore, our extensive experimental tests have led us to some interesting conclusions. We found out that D-Wave's {\tt Advantage\_system4.1} is more efficient than {\tt Advantage\_system1.1} both in terms of qubit utilization and quality of solutions. Using our proposed matrix formulation, it is possible to run the {\tt Burma14} instance of the TSPLIB library on a D-Wave's {\tt Advantage\_system4.1}, although without obtaining a ``correct'' solution. Finally, we have established experimentally that D-Wave's Hybrid solvers always provide a valid solution to a problem and never break the constraints of a QUBO model, even for arbitrarily big problems, of the order of $120$ nodes.

\subsection{Organization}

This paper is structured as follows. Section \ref{sec:Introduction} gives an introduction to the subject along with the most relevant references. Section \ref{sec:The Standard QUBO Formulation} introduces and explains the standard QUBO formulation of the HCP and TSP problems. Section \ref{sec:The Matrix QUBO Formulation}, through a detailed mathematical analysis, presents the matrix QUBO formulation of the HCP and TSP problems. Section \ref{sec:Normalizing Graph Weights} demonstrates the graph weight normalization procedure. Section \ref{sec:Experimenting on D-Wave's QPU} contains the experimental results obtained from our tests on D-Waves's QPU. Section \ref{sec:Experimenting on D-Wave's Leap’s Hybrid Solvers} presents the results from the tests contacted on D-Waves's Hybrid Solver, and section \ref{sec:Discussion and Conclusions} summarizes and discusses the conclusions drawn from the experiments.

\section{The standard QUBO formulation} \label{sec:The Standard QUBO Formulation}

We recall that given a graph $G = ( V, E )$, where $V$ is the set of vertices and $E$ is the set of edges, both HCP and TSP problems involve finding a tour

\begin{align} \label{eq:Tour}
	T = ( p_1, \dots, p_n, p_{ n + 1 } ) \ , \ p_1 = p_{ n + 1 } \ ,
\end{align}

where $n$ is the number of vertices, and $p_i$ is the vertex in the $i^{th}$ position of the tour.

Typically, QUBO expressions are formed from binary decision variables, i.e., variables that can only take the values $0$ or $1$. In this context, these binary variables will be designated by $x_{v, p}$ and they will have the usual meaning:

\begin{align} \label{eq:Binary Decision Variables}
	x_{v, p}
	=
	\left\{
	\begin{matrix*}[l]
		1 & \text{vertex } v \text{ is at position } p \text{ in the tour} \ , \\
		0 & \text{otherwise} \ .
	\end{matrix*}
	\right.
\end{align}

The HCP and TSP QUBO forms, as given by \cite{Lucas2014}, are built-up by the following simpler Hamiltonians.

\begin{center}
	\begin{minipage}{0.45 \textwidth}
		\begin{align}
			H_{v, p} = \sum_{v = 1}^{n} \left( 1 - \sum_{p = 1}^{n} x_{v, p} \right)^{2} \label{eq:Each Vertex Appears in Exactly One Position Hamiltonian}
		\end{align}
	\end{minipage}
	\begin{minipage}{0.45 \textwidth}
		\begin{align}
			H_{p, v} = \sum_{p = 1}^{n} \left( 1 - \sum_{v = 1}^{n} x_{v, p} \right)^{2} \label{eq:Each Position is Occupied by Exactly One Vertex Hamiltonian}
		\end{align}
	\end{minipage}

	\begin{minipage}{0.45 \textwidth}
		\begin{align}
			H_{E^c} = \sum_{(u, v) \not \in E}^{} \ \sum_{p = 1}^{n} x_{u, p} x_{v, p + 1} \label{eq:No Edge in the Graph Complement Hamiltonian}
		\end{align}
	\end{minipage}
	\begin{minipage}{0.45 \textwidth}
		\begin{align}
			H_{W} = \sum_{(u, v) \in E}^{} w_{u, v} \sum_{p = 1}^{n} x_{u, p} x_{v, p + 1} \label{eq:Tour Cost Hamiltonian}
		\end{align}
	\end{minipage}
\end{center}

These $4$ Hamiltonians, $H_{v, p}, H_{p, v}, H_{E^c}$ and $H_{W}$, encode fundamental properties that capture the essence of the two problems.

\begin{enumerate}
	\item	$H_{v, p}$ asserts that every vertex must appears in exactly one position of the tour.
	\item	$H_{p, v}$ states that each position of the tour is occupied by precisely one vertex.
	\item	$H_{E^c}$ requires that the tour is comprised of edges that ``really'' exist. If the tour, mistakenly, contains a ``phantom'' edge, i.e., an edge belonging to $E^c$, then this will incur an energy penalty.
	\item	$H_{W}$ computes the cost or weight of the tour. A tour minimizing this Hamiltonian is an optimal tour.
\end{enumerate}

The complete QUBO forms of the HCP and TSP problems are given below.

\begin{center}
	\begin{minipage}{0.45 \textwidth}
		\begin{align}
			H_{HCP} = c_1 \left( H_{v, p} + H_{p, v} + H_{E^c} \right) \label{eq:HCP QUBO}
		\end{align}
	\end{minipage}
	\begin{minipage}{0.45 \textwidth}
		\begin{align}
			H_{TSP} = H_{HCP} + c_2 H_{W} \label{eq:TSP QUBO}
		\end{align}
	\end{minipage}
\end{center}

In the above formulas $c_1$ and $c_2$ are positive constants, different, in general. We shall have more to say later about the significance of these constants. The first $3$ Hamiltonians, $H_{v, p}, H_{p, v}, H_{E^c}$, express integrity constraints; the slightest violations renders the solution invalid. The fourth Hamiltonian, $H_{W}$, is the minimization objective, that is necessary in order to converge to the tour associated with the minimal cost.

\begin{figure}[H]
	\begin{tcolorbox}
		[
			grow to left by = 0.0 cm,
			grow to right by = 0.0 cm,
			colback = gray!03,
			enhanced jigsaw, 
			sharp corners,
			toprule = 1.0 pt,
			bottomrule = 1.0 pt,
			leftrule = 0.1 pt,
			rightrule = 0.1 pt,
			sharp corners,
			center title,
			fonttitle = \bfseries
		]
		\begin{center}
			\begin{minipage}[b]{0.45 \textwidth}
				\centering
				\begin{tikzpicture} [ scale = 1.0 ]
					\def \angle {360 / 8}
					\node [ circle, fill = WordAquaLighter60, minimum size = 7 mm ] ( N1 ) at ( { 3 * cos(1 * \angle) }, { 3 * sin(1 * \angle) } ) { $\mathbf{1}$ };
					\node [ circle, fill = WordAquaLighter60, minimum size = 7 mm ] ( N2 ) at ( { 3 * cos(3 * \angle) }, { 3 * sin(3 * \angle) } ) { $\mathbf{2}$ };
					\node [ circle, fill = WordAquaLighter60, minimum size = 7 mm ] ( N3 ) at ( { 3 * cos(5 * \angle) }, { 3 * sin(5 * \angle) } ) { $\mathbf{3}$ };
					\node [ circle, fill = WordAquaLighter60, minimum size = 7 mm ] ( N4 ) at ( { 3 * cos(7 * \angle) }, { 3 * sin(7 * \angle) } ) { $\mathbf{4}$ };
					\draw [ -, >=stealth, line width = 0.25 mm, shorten >= 3.5 mm, shorten <= 3.5 mm ] ( { 3 * cos(1 * \angle) }, { 3 * sin(1 * \angle) } ) -- ( { 3 * cos(3 * \angle) }, { 3* sin(3 * \angle) } ) node [ midway, above ] (W12) {$30$};
					\draw [ -, >=stealth, line width = 0.25 mm, shorten >= 3.5 mm, shorten <= 3.5 mm ] ( { 3 * cos(3 * \angle) }, { 3 * sin(3 * \angle) } ) -- ( { 3 * cos(5 * \angle) }, { 3* sin(5 * \angle) } ) node [ midway, left ] (W23) {$20$};
					\draw [ -, >=stealth, line width = 0.25 mm, shorten >= 3.5 mm, shorten <= 3.5 mm ] ( { 3 * cos(5 * \angle) }, { 3 * sin(5 * \angle) } ) -- ( { 3 * cos(7 * \angle) }, { 3* sin(7 * \angle) } ) node [ midway, below ] (W34) {$35$};
					\draw [ -, >=stealth, line width = 0.25 mm, shorten >= 3.5 mm, shorten <= 3.5 mm ] ( { 3 * cos(7 * \angle) }, { 3 * sin(7 * \angle) } ) -- ( { 3 * cos(1 * \angle) }, { 3* sin(1 * \angle) } ) node [ midway, right ] (W41) {$12$};
					\draw [ -, >=stealth, line width = 0.25 mm, shorten >= 3.5 mm, shorten <= 3.5 mm ] ( { 3 * cos(1 * \angle) }, { 3 * sin(1 * \angle) } ) -- ( { 3 * cos(5 * \angle) }, { 3* sin(5 * \angle) } ) node [ midway, xshift = 0.3 cm, yshift = 0.7 cm ] (W13) {$42$};
					\draw [ -, >=stealth, line width = 0.25 mm, shorten >= 3.5 mm, shorten <= 3.5 mm ] ( { 3 * cos(3 * \angle) }, { 3 * sin(3 * \angle) } ) -- ( { 3 * cos(7 * \angle) }, { 3* sin(7 * \angle) } ) node [ midway, xshift = 0.3 cm, yshift = -0.7 cm ] (W24) {$34$};
				\end{tikzpicture}
				\caption{The $4$-node graph $G_{1}$ of Example \ref{xmp:Standard QUBO Example}.}
				\label{fig:4 Node Graph Example}
			\end{minipage}
			\hfill
			\begin{minipage}[b]{0.45 \textwidth}
				\begin{align} \label{eq:G_1 Adjacency Matrix}
					W
					=
					\begin{bNiceMatrix}
						0  & 30 & 42 & 12 \\
						30 &  0 & 20 & 34 \\
						42 & 20 &  0 & 35 \\
						12 & 34 & 35 &  0 \\
					\end{bNiceMatrix}
				\end{align}
				\caption{The adjacency matrix of the graph shown in Figure \ref{fig:4 Node Graph Example}.}
				\label{fig:4 Node Graph Adjacency Matrix}
			\end{minipage}
		\end{center}
	\end{tcolorbox}
\end{figure}

\begin{example} \label{xmp:Standard QUBO Example}
	To provide a clearer picture, we will give a small scale example of formulating a TSP into QUBO, using the graph $G_{1}$ shown in Figure \ref{fig:4 Node Graph Example}. The problem's graph has $n = 4$ nodes, with its adjacency matrix being Figure \ref{fig:4 Node Graph Adjacency Matrix}. Using equations (\ref{eq:Each Vertex Appears in Exactly One Position Hamiltonian})--(\ref{eq:TSP QUBO}) we can construct the Hamiltonian of the QUBO model for solving the TSP in this graph, as shown below. To emphasize the fact the Hamiltonians computed in this example refer to the particular graph $G_{1}$, we use the notation $H_{v, p}^{ G_{1} }$, $H_{p, v}^{ G_{1} }$, $H_{E^c}^{ G_{1} }$, $H_{W}^{ G_{1} }$, $H_{HCP}^{ G_{1} }$ and $H_{TSP}^{ G_{1} }$ when referring to them.

	In this example, the general expression $\left( 1 - \sum_{p = 1}^{n} x_{v, p} \right)^{2}$ reduces to $\left( 1 - x_{v, 1} - x_{v, 2} - x_{v, 3} - x_{v, 4} \right)^{2}$. This expression can be further simplified by taking into account the identity
	\begin{align} \label{eq:Sqaure of Polynomial}
		&\hspace{0.4 cm} ( 1 - a - b - c - d )^2
		\nonumber \\
		&= 1 + a^2 + b^2 + c^2 + d^2 - 2 a - 2 b - 2 c - 2 d + 2 a b + 2 a c + 2 a d + 2 b c + 2 b d + 2 c d
	\end{align}
	and the fact that the square of a binary decision variable is just the variable
	\begin{align} \label{eq:Sqaure of Binary Variables}
		x_{v, p}^2 = x_{v, p} \ ,
	\end{align}
	to give
	\begin{align} \label{eq:Square Constraint Identity}
		\left( 1 - \sum_{p = 1}^{4} x_{v, p} \right)^{2}
		&=
		\left( 1 - x_{v, 1} - x_{v, 2} - x_{v, 3} - x_{v, 4} \right)^{2}
		\nonumber \\
		&=
		1 + x_{v, 1} + x_{v, 2} + x_{v, 3} + x_{v, 4} - 2 x_{v, 1} - 2 x_{v, 2} - 2 x_{v, 3} - 2 x_{v, 4}
		\nonumber \\
		&+ 2 x_{v, 1} x_{v, 2} + 2 x_{v, 1} x_{v, 3} + 2 x_{v, 1} x_{v, 4} + 2 x_{v, 2} x_{v, 3} + 2 x_{v, 2} x_{v, 4} + 2 x_{v, 3} x_{v, 4}
		\nonumber \\
		&=
		1 - x_{v, 1} - x_{v, 2} - x_{v, 3} - x_{v, 4}
		\nonumber \\
		&+ 2 x_{v, 1} x_{v, 2} + 2 x_{v, 1} x_{v, 3} + 2 x_{v, 1} x_{v, 4} + 2 x_{v, 2} x_{v, 3} + 2 x_{v, 2} x_{v, 4} + 2 x_{v, 3} x_{v, 4}
	\end{align}
	Therefore,
	{\small
		\begin{align} \label{eq:H_v_p Constraints}
			H_{v, p}^{ G_{1} } = \sum_{v = 1}^{4} \left( 1 - \sum_{p = 1}^{4} x_{v, p} \right)^{2}
			\nonumber \\
			=
			4
			- x_{1, 1} - x_{1, 2} - x_{1, 3} - x_{1, 4}
			- x_{2, 1} - x_{2, 2} - x_{2, 3} - x_{2, 4}
			\nonumber \\
			- x_{3, 1} - x_{3, 2} - x_{3, 3} - x_{3, 4}
			- x_{4, 1} - x_{4, 2} - x_{4, 3} - x_{4, 4}
			\nonumber \\
			+ 2 x_{1, 1} x_{1, 2} + 2 x_{1, 1} x_{1, 3} + 2 x_{1, 1} x_{1, 4} + 2 x_{1, 2} x_{1, 3} + 2 x_{1, 2} x_{1, 4} + 2 x_{1, 3} x_{1, 4}
			\nonumber \\
			+ 2 x_{2, 1} x_{2, 2} + 2 x_{2, 1} x_{2, 3} + 2 x_{2, 1} x_{2, 4} + 2 x_{2, 2} x_{2, 3} + 2 x_{2, 2} x_{2, 4} + 2 x_{2, 3} x_{2, 4}
			\nonumber \\
			+ 2 x_{3, 1} x_{3, 2} + 2 x_{3, 1} x_{3, 3} + 2 x_{3, 1} x_{3, 4} + 2 x_{3, 2} x_{3, 3} + 2 x_{3, 2} x_{3, 4} + 2 x_{3, 3} x_{3, 4}
			\nonumber \\
			+ 2 x_{4, 1} x_{4, 2} + 2 x_{4, 1} x_{4, 3} + 2 x_{4, 1} x_{4, 4} + 2 x_{4, 2} x_{4, 3} + 2 x_{4, 2} x_{4, 4} + 2 x_{4, 3} x_{4, 4}
		\end{align}
	}
	In a similar way, we may construct $H_{p, v}^{ G_{1} }$.
	{\small
		\begin{align} \label{eq:H_p_v Constraints}
			H_{p, v}^{ G_{1} } = \sum_{p = 1}^{4} \left( 1 - \sum_{v = 1}^{4} x_{v, p} \right)^{2}
			\nonumber \\
			=
			4
			- x_{1, 1} - x_{2, 1} - x_{3, 1} - x_{4, 1}
			- x_{1, 2} - x_{2, 2} - x_{3, 2} - x_{4, 2}
			\nonumber \\
			- x_{1, 3} - x_{2, 3} - x_{3, 3} - x_{4, 3}
			- x_{1, 4} - x_{2, 4} - x_{3, 4} - x_{4, 4}
			\nonumber \\
			+ 2 x_{1, 1} x_{2, 1} + 2 x_{1, 1} x_{3, 1} + 2 x_{1, 1} x_{4, 1} + 2 x_{2, 1} x_{3, 1} + 2 x_{2, 1} x_{4, 1} + 2 x_{3, 1} x_{4, 1}
			\nonumber \\
			+ 2 x_{1, 2} x_{2, 2} + 2 x_{1, 2} x_{3, 2} + 2 x_{1, 2} x_{4, 2} + 2 x_{2, 2} x_{3, 2} + 2 x_{2, 2} x_{4, 2} + 2 x_{3, 2} x_{4, 2}
			\nonumber \\
			+ 2 x_{1, 3} x_{2, 3} + 2 x_{1, 3} x_{3, 3} + 2 x_{1, 3} x_{4, 3} + 2 x_{2, 3} x_{3, 3} + 2 x_{2, 3} x_{4, 3} + 2 x_{3, 3} x_{4, 3}
			\nonumber \\
			+ 2 x_{1, 4} x_{2, 4} + 2 x_{1, 4} x_{3, 4} + 2 x_{1, 4} x_{4, 4} + 2 x_{2, 4} x_{3, 4} + 2 x_{2, 4} x_{4, 4} + 2 x_{3, 4} x_{4, 4}
		\end{align}
	}
	Each of the expansions of $H_{v, p}^{ G_{1} }$ and $H_{p, v}^{ G_{1} }$, as given in (\ref{eq:H_v_p Constraints}) and (\ref{eq:H_p_v Constraints}), creates a constant term, namely $4$, and $40$ constraints in total: $16$ linear constraints and $24$ quadratic constraints. The linear constraints are the \emph{same}, but the quadratic constraints are \emph{different}.

	$H_{E^c}$ depends on the topology of the graph. In our example, where the graph is complete, $H_{E^c}^{ G_{1} }$ would only involve self-loops, that is
	{\small
		\begin{align} \label{eq:H_E^c Constraints for Complete Graph}
			H_{E^c}^{ G_{1} } = \sum_{(u, v) \not \in E}^{} \ \sum_{p = 1}^{4} x_{u, p} x_{v, p + 1}
			\nonumber \\
			=
			x_{1, 1} x_{1, 2} + x_{1, 2} x_{1, 3} + x_{1, 3} x_{1, 4} + x_{1, 4} x_{1, 1}
			+
			x_{2, 1} x_{2, 2} + x_{2, 2} x_{2, 3} + x_{2, 3} x_{2, 4} + x_{2, 4} x_{2, 1}
			\nonumber \\
			+
			x_{3, 1} x_{3, 2} + x_{3, 2} x_{3, 3} + x_{3, 3} x_{3, 4} + x_{3, 4} x_{3, 1}
			+
			x_{4, 1} x_{4, 2} + x_{4, 2} x_{4, 3} + x_{4, 3} x_{4, 4} + x_{4, 4} x_{4, 1}
		\end{align}
	}
	To avoid any potential confusion, let us point out that in the tour, which is a cycle, position $n + 1$ coincides with position $1$, a fact which was used in the formulation of (\ref{eq:H_E^c Constraints for Complete Graph}). As a result of the completeness of the graph, all the above $16$ quadratic constraints are \emph{already present} in $H_{v, p}^{ G_{1} }$, as can be verified by recalling (\ref{eq:H_v_p Constraints}). Of course, their presence in $H_{E^c}^{ G_{1} }$ increases by $1$ their corresponding coefficient, amplifying their relative significance. It is useful to briefly consider what happens when the graph is not complete, i.e., at least one edge that is not a self-loop is missing. If, for instance, $G = ( V, E )$ is an \emph{undirected} graph and $(1, 3) \not \in E$, which implies that also $(3, 1) \not \in E$, then $H_{E^c}^{ G_{1} }$, apart from the constraints shown in (\ref{eq:H_E^c Constraints for Complete Graph}), would additionally, contain the following constraints.
	{\small
		\begin{align} \label{eq:H_E^c Constraints for Incomplete Undirected Graph}
			x_{1, 1} x_{3, 2} + x_{1, 2} x_{3, 3} + x_{1, 3} x_{3, 4} + x_{1, 4} x_{3, 1}
			+
			x_{3, 1} x_{1, 2} + x_{3, 2} x_{1, 3} + x_{3, 3} x_{1, 4} + x_{3, 4} x_{1, 1}
		\end{align}
	}
	If, on the other hand, $G = ( V, E )$ is a \emph{directed} graph and $(1, 3) \not \in E$, which does not necessarily imply that $(3, 1) \not \in E$, then, $H_{E^c}^{ G_{1} }$, apart from the constraints shown in (\ref{eq:H_E^c Constraints for Complete Graph}), would additionally, contain the following constraints.
	{\small
		\begin{align} \label{eq:H_E^c Constraints for Incomplete Directed Graph}
			x_{1, 1} x_{3, 2} + x_{1, 2} x_{3, 3} + x_{1, 3} x_{3, 4} + x_{1, 4} x_{3, 1}
		\end{align}
	}

	$H_{W}$ captures the topology of the graph and the cost, also referred to as weight, associated with each edge. In this particular example, it will produce the constrains listed below, where we have again identified position $n + 1$ with position $1$.
	\begin{tcolorbox}
		[
			grow to left by = 0.75 cm,
			grow to right by = 0.75 cm,
			colback = white, 
			enhanced jigsaw, 
			sharp corners,
			boxrule = 0.1 pt,
			toprule = 0.1 pt,
			bottomrule = 0.1 pt
		]
		{\small
		\begin{align} \label{eq:H_W Constraints for Complete Graph}
			H_{W}^{ G_{1} }
			&=
			\sum_{(u, v) \in E}^{} w_{u, v} \sum_{p = 1}^{4} x_{u, p} x_{v, p + 1}
			\nonumber \\
			&=
			30 x_{1, 1} x_{2, 2} + 30 x_{1, 2} x_{2, 3} + 30 x_{1, 3} x_{2, 4} + 30 x_{1, 4} x_{2, 1}
			+
			42 x_{1, 1} x_{3, 2} + 42 x_{1, 2} x_{3, 3} + 42 x_{1, 3} x_{3, 4} + 42 x_{1, 4} x_{3, 1}
			\nonumber \\
			&+
			12 x_{1, 1} x_{4, 2} + 12 x_{1, 2} x_{4, 3} + 12 x_{1, 3} x_{4, 4} + 12 x_{1, 4} x_{4, 1}
			+
			30 x_{2, 1} x_{1, 2} + 30 x_{2, 2} x_{1, 3} + 30 x_{2, 3} x_{1, 4} + 30 x_{2, 4} x_{1, 1}
			\nonumber \\
			&+
			20 x_{2, 1} x_{3, 2} + 20 x_{2, 2} x_{3, 3} + 20 x_{2, 3} x_{3, 4} + 20 x_{2, 4} x_{3, 1}
			+
			34 x_{2, 1} x_{4, 2} + 34 x_{2, 2} x_{4, 3} + 34 x_{2, 3} x_{4, 4} + 34 x_{2, 4} x_{4, 1}
			\nonumber \\
			&+
			42 x_{3, 1} x_{1, 2} + 42 x_{3, 2} x_{1, 3} + 42 x_{3, 3} x_{1, 4} + 42 x_{3, 4} x_{1, 1}
			+
			20 x_{3, 1} x_{2, 2} + 20 x_{3, 2} x_{2, 3} + 20 x_{3, 3} x_{2, 4} + 20 x_{3, 4} x_{2, 1}
			\nonumber \\
			&+
			35 x_{3, 1} x_{4, 2} + 35 x_{3, 2} x_{4, 3} + 35 x_{3, 3} x_{4, 4} + 35 x_{3, 4} x_{4, 1}
			+
			12 x_{4, 1} x_{1, 2} + 12 x_{4, 2} x_{1, 3} + 12 x_{4, 3} x_{1, 4} + 12 x_{4, 4} x_{1, 1}
			\nonumber \\
			&+
			34 x_{4, 1} x_{2, 2} + 34 x_{4, 2} x_{2, 3} + 34 x_{4, 3} x_{2, 4} + 34 x_{4, 4} x_{2, 1}
			+
			35 x_{4, 1} x_{3, 2} + 35 x_{4, 2} x_{3, 3} + 35 x_{4, 3} x_{3, 4} + 35 x_{4, 4} x_{3, 1}
		\end{align}
		}
	\end{tcolorbox}
	The expansion of $H_{W}^{ G_{1} }$ shown in (\ref{eq:H_W Constraints for Complete Graph}) creates $48$ \emph{new} quadratic constrains. It is instructive to see what changes if the graph is not complete, assuming, as before, that the missing edges are not self-loops. If the are $m$ edges present, then each such edge will involve $n$ (here $n = 4$ quadratic constraints).

	In view of (\ref{eq:HCP QUBO}), the Hamiltonian for solving the HCP on $G_{1}$ is $H_{HCP}^{ G_{1} } = c_1 \left( H_{v, p}^{ G_{1} } + H_{p, v}^{ G_{1} } + H_{E^c}^{ G_{1} } \right)$, where $c_1$ is a positive constant. $H_{HCP}^{ G_{1} }$ is constructed by combining (\ref{eq:H_v_p Constraints}) and (\ref{eq:H_p_v Constraints}), it requires $16$ binary variables, and it involves $64$ constraints. Analogously, the Hamiltonian for solving the TSP on $G_{1}$ is $H_{TSP}^{ G_{1} } = H_{HCP}^{ G_{1} } + c_2 H_{W}^{ G_{1} }$, where $c_2$ is also a positive constant. Since $H_{TSP}^{ G_{1} }$ is constructed by combining (\ref{eq:H_v_p Constraints}), (\ref{eq:H_p_v Constraints}) and (\ref{eq:H_W Constraints for Complete Graph}), it contains $112$ constraints.
	\hfill $\triangleleft$
\end{example}

Using the previous example as a starting point, it is straightforward to derive the number of constraints that for a general graph with $n$ nodes. At this point, it is expedient to make the following remark. In the QUBO formulation, the binary variables corresponding to an edge (or arc) $( u, v )$ are distinct from the binary variables corresponding to the edge (or arc) $( v, u )$. Therefore, they give rise to distinct constraints in the Hamiltonians $H_{HCP}$ and $H_{TSP}$. Hence, from this perspective, irrespective of whether a complete graph $G = ( V, E )$ with $n$ nodes is directed or undirected, it should be considered as having $n ( n - 1 )$ edges. In other words, even in a undirected graph, we should view the edges $( u, v )$ and $( v, u )$ as different.

\begin{itemize}
	\item	Each of the Hamiltonians $H_{v, p}$ and $H_{p, v}$, as given in (\ref{eq:H_v_p Constraints}) and (\ref{eq:H_p_v Constraints}), require a constant term, namely $n$, $n^{2}$ linear constraints, and $\frac{ n^{2} ( n - 1 ) }{2}$ quadratic constraints, which brings the total number of constraints to $\frac{ n^{2} ( n + 1 ) }{2}$. It important to clarify that the linear constraints are the same in both $H_{v, p}$ and $H_{p, v}$, but the quadratic constraints are different.
	\item	If the graph is complete, then $H_{E^c}$ does not add any new constraints. To be precise, $H_{E^c}$ involves $n^{2}$ quadratic constraints, which are also present in the $H_{v, p}$ Hamiltonian. Hence, $H_{E^c}$ does not add new constraints, it only enhances the relative weight of existing constraints.
	\item	If the graph is not complete, then $H_{E^c}$ creates for each \emph{missing} edge $(u, v) \neq E$ that is not a self-loop, i.e., $u \neq v$, $n$ additional constraints. If $k$ such edges are \emph{missing}, there will be $n k$ additional quadratic constraints in total. This fact is experimentally validated from the results presented in subsection \ref{subsec:The Effect of Graph Connectivity on Qubit Usage}.
	\item	For a complete graph with $n$ nodes $H_{W}$ introduces $n^{2} ( n - 1 )$ quadratic constraints. One could, equivalently, assert that in a complete graph $H_{W}$ requires $n | E |$ quadratic constraints, where $| E |$ stands for the number of edges in the graph. This last formula can be generalized further to the case where the graph is not complete and has $m$ edges. In such a case $H_{W}$ will add $n m$ quadratic constraints.
	\item	For a complete graph, $H_{HCP}$ will require $n^{2}$ binary variables and $n^{3}$ constraints in total (recall that the $n^{2}$ linear constraints are the same). If the graph is not complete, it will require even more constraints, specifically $n^{3} + n k$, where $k$ is the number of \emph{missing} edges.
	\item	For a complete graph, $H_{TSP}$ will require $n^{2}$ binary variables, $2 n^{2} ( n - 1 ) = 2 n | E |$ quadratic constraints, and $n^{2} ( 2 n - 1 )$ constraints in total. If the graph is not complete, it will require $n^{3} + n ( m + k )$, where $m$ is the number of \emph{existing} edges and $k$ is the number of \emph{missing} edges.
	\item	In the typical and practically important case where we have a \emph{simple} graph, that is a graph with no \emph{self-loops} and no \emph{parallel} edges, then $m + k =n( n - 1 )$. This leads to the simplification of the previous formulas, which now become $2 n^{2} ( n - 1 )$ for the number of quadratic constraints and $n^{2} ( 2 n - 1 )$ for the total number of constraints. Let us point out the interesting fact that in this case the number of constraints required for the $H_{TSP}$ Hamiltonian is equal to the number of constraints required when the graph is complete.
\end{itemize}

Tables \ref{tbl:Number of Constraints in Each Hamiltonian for a Complete Graph}, \ref{tbl:Number of Constraints in Each Hamiltonian for an Arbitrary Graph}, and \ref{tbl:Number of Constraints in Each Hamiltonian for a Simple Graph without Self-Loops} summarize the above observations regarding the number of constraints involved in the HCP and TSP Hamiltonians.

\begin{table}[H]
	\renewcommand{\arraystretch}{2.0}
	\begin{tcolorbox}
		[
			grow to left by = 0.00 cm,
			grow to right by = 0.00 cm,
			colback = gray!03,
			enhanced jigsaw, 
			sharp corners,
			boxrule = 0.1 pt,
			toprule = 0.1 pt,
			bottomrule = 0.1 pt
		]
		\caption{This table summarizes the number of constraints that are involved in each of the Hamiltonians (\ref{eq:Each Vertex Appears in Exactly One Position Hamiltonian})--(\ref{eq:TSP QUBO}), when applied to a \textbf{complete} graph with $n$ nodes. \\}
		\label{tbl:Number of Constraints in Each Hamiltonian for a Complete Graph}
		\centering
		\begin{tabular}
			{
				>{\centering\arraybackslash} m{2.50 cm} !{\vrule width 1.25 pt}
				>{\centering\arraybackslash} m{1.50 cm} !{\vrule width 0.5 pt}
				>{\centering\arraybackslash} m{3.00 cm} !{\vrule width 0.5 pt}
				>{\centering\arraybackslash} m{3.00 cm}
			}
			\Xhline{4\arrayrulewidth}
			\multicolumn{4}{c}{\# \textbf{Binary variables:} $n^{2}$}
			\\
			&
			\multicolumn{3}{c}{\# \textbf{Constraints}} 
			\\
			\Xhline{\arrayrulewidth}
			\textbf{Hamiltonian}
			&
			Linear
			&
			Quadratic
			&
			Total
			\\
			\Xhline{3\arrayrulewidth}
			$H_{v, p}$
			&
			$n^{2}$
			&
			$\frac{ n^{2} ( n - 1 ) }{2} = n | E |$
			&
			$\frac{ n^{2} ( n + 1 ) }{2}$
			\\
			\Xhline{\arrayrulewidth}
			$H_{p, v}$
			&
			$n^{2}$
			&
			$\frac{ n^{2} ( n - 1 ) }{2} = n | E |$
			&
			$\frac{ n^{2} ( n + 1 ) }{2}$
			\\
			\Xhline{\arrayrulewidth}
			$H_{E^c}$
			&
			$0$
			&
			$0$
			&
			$0$
			\\
			\Xhline{\arrayrulewidth}
			$H_{W}$
			&
			$0$
			&
			$n^{2} ( n - 1 ) = n | E |$
			&
			$n^{2} ( n - 1 ) = n | E |$
			\\
			\Xhline{\arrayrulewidth}
			$H_{HCP}$
			&
			$n^{2}$
			&
			$n^{2} ( n - 1 )$
			&
			$n^{3}$
			\\
			\Xhline{\arrayrulewidth}
			$H_{TSP}$
			&
			$n^{2}$
			&
			$2 n^{2} ( n - 1 ) = 2 n | E |$
			&
			$n^{2} ( 2 n - 1 )$
			\\
			\Xhline{4\arrayrulewidth}
		\end{tabular}
	\end{tcolorbox}
	\renewcommand{\arraystretch}{1.0}
\end{table}

\begin{table}[H]
	\renewcommand{\arraystretch}{2.0}
	\begin{tcolorbox}
		[
			grow to left by = 0.00 cm,
			grow to right by = 0.00 cm,
			colback = gray!03,
			enhanced jigsaw, 
			sharp corners,
			boxrule = 0.1 pt,
			toprule = 0.1 pt,
			bottomrule = 0.1 pt
		]
		\caption{This table summarizes the number of constraints that are involved in each of the Hamiltonians (\ref{eq:Each Vertex Appears in Exactly One Position Hamiltonian})--(\ref{eq:TSP QUBO}), when applied to an arbitrary graph with $n$ nodes, $m$ edges and $k$ missing edges (none of which is a self-loop). \\}
		\label{tbl:Number of Constraints in Each Hamiltonian for an Arbitrary Graph}
		\centering
		\begin{tabular}
			{
				>{\centering\arraybackslash} m{2.50 cm} !{\vrule width 1.25 pt}
				>{\centering\arraybackslash} m{1.50 cm} !{\vrule width 0.5 pt}
				>{\centering\arraybackslash} m{3.00 cm} !{\vrule width 0.5 pt}
				>{\centering\arraybackslash} m{3.00 cm}
			}
			\Xhline{4\arrayrulewidth}
			\multicolumn{4}{c}{\# \textbf{Binary variables:} $n^{2}$}
			\\
			&
			\multicolumn{3}{c}{\# \textbf{Constraints}} 
			\\
			\Xhline{\arrayrulewidth}
			\textbf{Hamiltonian}
			&
			Linear
			&
			Quadratic
			&
			Total
			\\
			\Xhline{3\arrayrulewidth}
			$H_{v, p}$
			&
			$n^{2}$
			&
			$\frac{ n^{2} ( n - 1 ) }{2}$
			&
			$\frac{ n^{2} ( n + 1 ) }{2}$
			\\
			\Xhline{\arrayrulewidth}
			$H_{p, v}$
			&
			$n^{2}$
			&
			$\frac{ n^{2} ( n - 1 ) }{2}$
			&
			$\frac{ n^{2} ( n + 1 ) }{2}$
			\\
			\Xhline{\arrayrulewidth}
			$H_{E^c}$
			&
			$0$
			&
			$n k$
			&
			$n k$
			\\
			\Xhline{\arrayrulewidth}
			$H_{W}$
			&
			$0$
			&
			$n m$
			&
			$n m$
			\\
			\Xhline{\arrayrulewidth}
			$H_{HCP}$
			&
			$n^{2}$
			&
			$n^{2} ( n - 1 ) + n k$
			&
			$n^{3} + n k$
			\\
			\Xhline{\arrayrulewidth}
			$H_{TSP}$
			&
			$n^{2}$
			&
			$n^{2} ( n - 1 ) + n ( m + k )$
			&
			$n^{3} + n ( m + k )$
			\\
			\Xhline{4\arrayrulewidth}
		\end{tabular}
	\end{tcolorbox}
	\renewcommand{\arraystretch}{1.0}
\end{table}

\begin{table}[H]
	\renewcommand{\arraystretch}{2.0}
	\begin{tcolorbox}
		[
			grow to left by = 0.00 cm,
			grow to right by = 0.00 cm,
			colback = gray!03,
			enhanced jigsaw, 
			sharp corners,
			boxrule = 0.1 pt,
			toprule = 0.1 pt,
			bottomrule = 0.1 pt
		]
		\caption{This table shows the number of constraints that are involved in the $H_{TSP}$ Hamiltonian (\ref{eq:TSP QUBO}), when $m + k = n( n - 1 )$. \\}
		\label{tbl:Number of Constraints in Each Hamiltonian for a Simple Graph without Self-Loops}
		\centering
		\begin{tabular}
			{
				>{\centering\arraybackslash} m{2.50 cm} !{\vrule width 1.25 pt}
				>{\centering\arraybackslash} m{1.50 cm} !{\vrule width 0.5 pt}
				>{\centering\arraybackslash} m{3.00 cm} !{\vrule width 0.5 pt}
				>{\centering\arraybackslash} m{3.00 cm}
			}
			\Xhline{4\arrayrulewidth}
			\multicolumn{4}{c}{\# \textbf{Constraints} when $m + k = n( n - 1 )$}
			\\
			\Xhline{\arrayrulewidth}
			\textbf{Hamiltonian}
			&
			Linear
			&
			Quadratic
			&
			Total
			\\
			\Xhline{3\arrayrulewidth}

			$H_{TSP}$
			&
			$n^{2}$
			&
			$2 n^{2} ( n - 1 ) = 2 n | E |$
			&
			$n^{2} ( 2 n - 1 )$
			\\
			\Xhline{4\arrayrulewidth}
		\end{tabular}
	\end{tcolorbox}
	\renewcommand{\arraystretch}{1.0}
\end{table}

\section{The matrix QUBO formulation} \label{sec:The Matrix QUBO Formulation}

In this section we shall present the matrix form of the Hamiltonians (\ref{eq:Each Vertex Appears in Exactly One Position Hamiltonian})--(\ref{eq:TSP QUBO}) so as to enable their seamless execution of HCP and TSP instances by the quantum annealer and by tools like \texttt{qubovert} \cite{Qubovert2022}. For this purpose, we define for each $v \in V$ the column vector $X_{v}$ containing the binary decision variables $x_{v, 1}, x_{v, 2}, \dots, x_{v, n}$ corresponding to the possible positions of vertex $v$ in the tour. Then, by combining the $n$ vectors $X_{v}, 1 \leq v \leq n$, we construct the column vector $X$, which of course has $n^2$ rows.

\begin{tcolorbox}
	[
		grow to left by = - 0.0 cm,
		grow to right by = - 0.0 cm,
		colback = white, 
		enhanced jigsaw, 
		sharp corners,
		boxrule = 0.01 pt,
		toprule = 0.01 pt,
		bottomrule = 0.01 pt
	]
	\begin{center}
		\begin{minipage}{0.4 \textwidth}
			\begin{align} \label{eq:n-Dimensional Vector X_v}
				\NiceMatrixOptions{ xdots/shorten = 0.85 em, code-for-first-row = \color{GreenLighter2}, code-for-first-col = \color{RedPurple} }
				X_{v}
				= \quad
				\begin{bNiceMatrix}[ first-col ]
					& x_{v, 1}
					\\
					& x_{v, 2}
					\\
					\Vdots[ line-style = {solid, <->} ]_{n \text{ rows}} 
					& \vdots
					\\
					& x_{v, n}
				\end{bNiceMatrix}
			\end{align}
		\end{minipage}
		\begin{minipage}{0.4 \textwidth}
			\begin{align} \label{eq:n^2-Dimensional Vector X}
				\NiceMatrixOptions{ xdots/shorten = 0.85 em, code-for-first-row = \color{GreenLighter2}, code-for-last-col = \color{RedPurple} }
				X
				= \
				\begin{bNiceMatrix}[ margin, last-col, rules/width = 0.0025 mm ]
					X_{1} & \ \ \Vdots[ line-style = {solid, <->} ]^{n^{2} \text{ rows}} 
					\\
					\hline
					X_{2} &
					\\
					\hline
					\vdots &
					\\
					\hline
					X_{n} &
				\end{bNiceMatrix}
			\end{align}
		\end{minipage}
	\end{center}
\end{tcolorbox}

The matrices expressing (\ref{eq:HCP QUBO}) and (\ref{eq:TSP QUBO}) are based on the matrix form of (\ref{eq:Each Vertex Appears in Exactly One Position Hamiltonian})--(\ref{eq:Tour Cost Hamiltonian}). To define the latter in the easiest way, we need a few auxiliary $n \times n$ matrices. As usual, the $n \times n$ identity matrix is denoted by $I_{n}$, its negation by $-I_{n}$, and the $n \times n$ zero matrix whose entries are all $0$ by $O_{n}$.

\begin{tcolorbox}
	[
		grow to left by = 2.0 cm,
		grow to right by = 2.0 cm,
		colback = white, 
		enhanced jigsaw, 
		sharp corners,
		boxrule = 0.01 pt,
		toprule = 0.01 pt,
		bottomrule = 0.01 pt
	]
	\begin{center}
		\begin{minipage}{0.3 \textwidth}
			\begin{align} \label{eq:Identity Matrix I_n}
				\NiceMatrixOptions{ xdots/shorten = 0.85 em, code-for-first-row = \color{GreenLighter2}, code-for-first-col = \color{RedPurple} }
				I_{n}
				= \quad
				\begin{bNiceMatrix}[ right-margin = 1 mm, first-row, first-col ]
					& \Ldots[ line-style = {solid, <->}, shorten = 0pt ]^{n \text{ columns}} & \\
					\Vdots [ line-style = {solid, <->} ]_{n \text{ rows}} 
					& 1 & 0 & \Cdots & 0 \\
					& 0 & 1 & \Ddots & \Vdots \\
					& \Vdots & \Ddots & 1 & 0 \\
					& 0 & \Cdots & 0 & 1
					\CodeAfter \line{2-2}{3-3}
				\end{bNiceMatrix}
			\end{align}
		\end{minipage}
		\begin{minipage}{0.33 \textwidth}
			\begin{align} \label{eq:Identity Matrix - I_n}
				\NiceMatrixOptions{ xdots/shorten = 0.85 em, code-for-first-row = \color{GreenLighter2}, code-for-first-col = \color{RedPurple} }
				- I_{n}
				= \quad
				\begin{bNiceArray}{rrrr}[ right-margin = 1 mm, first-row, first-col ]
					& \Ldots[ line-style = {solid, <->}, shorten = 0pt ]^{n \text{ columns}} & \\
					\Vdots[ line-style = {solid, <->} ]_{n \text{ rows}} 
					& -1 & 0 & \Cdots & 0 \\
					& 0 & -1 & \Ddots & \Vdots \\
					& \Vdots & \Ddots & -1 & 0 \\
					& 0 & \Cdots & 0 & -1
					\CodeAfter \line{2-2}{3-3}
				\end{bNiceArray}
			\end{align}
		\end{minipage}
		\begin{minipage}{0.3 \textwidth}
			\begin{align} \label{eq:Zero Matrix O_n}
				\NiceMatrixOptions{ xdots/shorten = 0.85 em, code-for-first-row = \color{GreenLighter2}, code-for-first-col = \color{RedPurple} }
				O_{n}
				= \quad
				\begin{bNiceMatrix}[ right-margin = 1 mm, first-row, first-col ]
					& \Ldots[ line-style = {solid, <->}, shorten = 0pt ]^{n \text{ columns}} & \\
					\Vdots[ line-style = {solid, <->} ]_{n \text{ rows}} 
					& 0 & 0 & \Cdots & 0 \\
					& 0 & 0 & \Ddots & \Vdots \\
					& \Vdots & \Ddots & 0 & 0 \\
					& 0 & \Cdots & 0 & 0
					\CodeAfter \line{2-2}{3-3}
				\end{bNiceMatrix}
			\end{align}
		\end{minipage}
	\end{center}
\end{tcolorbox}

Two other auxiliary $n \times n$ matrices are $J_{n}$ and $K_{n}$ defined below.

\begin{tcolorbox}
	[
		grow to left by = 0.0 cm,
		grow to right by = 0.0 cm,
		colback = white, 
		enhanced jigsaw, 
		sharp corners,
		boxrule = 0.01 pt,
		toprule = 0.01 pt,
		bottomrule = 0.01 pt
	]
	\begin{center}
		\begin{minipage}{0.45 \textwidth}
			\begin{align} \label{eq:Auxiliary Matrix J_n}
				\NiceMatrixOptions{ xdots/shorten = 0.85 em, code-for-first-row = \color{GreenLighter2}, code-for-first-col = \color{RedPurple} }
				J_{n}
				= \quad
				\begin{bNiceArray}{rrrr}[ right-margin = 1 mm, first-row, first-col ]
					& \Ldots[ line-style = {solid, <->}, shorten = 0pt ]^{n \text{ columns}} & \\
					\Vdots[ line-style = {solid, <->} ]_{n \text{ rows}} 
					& - 1 & 2 & \Cdots & 2 \\
					& 0 & - 1 & \Ddots & \Vdots \\
					& \Vdots & \Ddots & - 1 & 2 \\
					& 0 & \Cdots & 0 & - 1
					\CodeAfter \line{2-2}{3-3}
				\end{bNiceArray}
			\end{align}
		\end{minipage}
		\begin{minipage}{0.475 \textwidth}
			\begin{align} \label{eq:Auxiliary Matrix K_n}
				\NiceMatrixOptions{ xdots/shorten = 0.85 em, code-for-first-row = \color{GreenLighter2}, code-for-first-col = \color{RedPurple} }
				K_{n}
				= \ \
				\begin{bNiceArray}{cccccc}[ margin = 1 mm, first-row, first-col ]
					& \Ldots[ line-style = {solid, <->}, shorten = 0pt ]^{n \text{ columns}} & \\
					\Vdots[ line-style = {solid, <->} ]_{n \text{ rows}} 
					& 0 & 1 & 0 & \Cdots & 0 & 1 \\
					& 0 & 0 & 1 & 0 & \Cdots & 0 \\
					& \Vdots & \Ddots & \Ddots & \Ddots & \Ddots & \Vdots\\
					& 0 & \Cdots & 0 & 0 & 1 & 0 \\
					& 0 & \Cdots & & & 0 & 1 \\
					& 0 & \Cdots & & & & 0
				\end{bNiceArray}
			\end{align}
		\end{minipage}
	\end{center}
\end{tcolorbox}

Let us assume that $W = [ w_{u, v} ]$ is the $n \times n$ adjacency matrix characterizing the topology of the given graph. In this matrix if $w_{u, v} = 0$ then the edge (or arc) $(u, v)$ does not exist in the graph, whereas if $w_{u, v} > 0$, then $w_{u, v}$ expresses the cost or weight associated with $(u, v)$. To each entry $w_{u, v}$ of $W$, we associate two $n \times n$ matrices $W_{(u, v)}$ and $W_{\overline{(u, v)}}$ defined as follows:


\begin{center}
	\begin{minipage}{0.45 \textwidth}
		\begin{align} \label{eq:Matrix W_(u, v)}
			W_{(u, v)} = w_{u, v} K_{n}
		\end{align}
	\end{minipage}
	\begin{minipage}{0.45 \textwidth}
		\begin{align} \label{eq:Matrix W_Co(u, v)}
			W_{\overline{(u, v)}} =
			\left\{
			\begin{matrix*}[l]
				K_{n}	&	\text{if } w_{u, v} = 0, \\
				O_{n}	&	\text{if } w_{u, v} > 0.
			\end{matrix*}
			\right.
		\end{align}
	\end{minipage}
\end{center}

We proceed now to give the matrix form of the Hamiltonians (\ref{eq:Each Vertex Appears in Exactly One Position Hamiltonian})--(\ref{eq:Tour Cost Hamiltonian}). Specifically we establish the correspondence between Hamiltonians and $n^2 \times n^2$ block matrices shown below, where the symbol $\mapsto$ is used to convey association and not equality.

\begin{tcolorbox}
	[
		grow to left by = 1.5 cm,
		grow to right by = 1.5 cm,
		colback = white, 
		enhanced jigsaw, 
		sharp corners,
		boxrule = 0.01 pt,
		toprule = 0.01 pt,
		bottomrule = 0.01 pt
	]
	\begin{center}
		\begin{minipage}{0.475 \textwidth}
			\begin{align} \label{eq:Matrix M_v_p}
				\NiceMatrixOptions{ xdots/shorten = 0.85 em, code-for-first-row = \color{GreenLighter2}, code-for-first-col = \color{RedPurple} }
				H_{v, p} \mapsto M_{v, p} = \quad
				\begin{bNiceArray}{cccc}[ margin = 1.2 mm, first-row, first-col, rules/color = WordAquaLighter40, rules/width = 0.2 mm ]
					& \Ldots[ line-style = {solid, <->}, shorten = 0pt ]^{n^2 \text{ columns}} & \\
					\Vdots[ line-style = {solid, <->} ]_{n^2 \text{ rows}} 
					& J_n & O_n & \Cdots & O_n \\
					& O_n & J_n & \Ddots & \Vdots \\
					& \Vdots & \Ddots & J_n & O_n \\
					& O_n & \Cdots & O_n & J_n
					\CodeAfter \line{2-2}{3-3}
				\end{bNiceArray}
			\end{align}
		\end{minipage}
		\begin{minipage}{0.475 \textwidth}
			\begin{align} \label{eq:Matrix M_p_v}
				\NiceMatrixOptions{ xdots/shorten = 0.85 em, code-for-first-row = \color{GreenLighter2}, code-for-first-col = \color{RedPurple} }
				H_{p, v} \mapsto M_{p, v} = \quad
				\begin{bNiceArray}{cccc}[ margin = 1.2 mm, first-row, first-col, rules/color = WordAquaLighter40, rules/width = 0.2 mm ]
					& \Ldots[ line-style = {solid, <->}, shorten = 0pt ]^{n^2 \text{ columns}} & \\
					\Vdots[ line-style = {solid, <->} ]_{n^2 \text{ rows}} 
					& - I_n & 2 \cdot I_n & \Cdots & 2 \cdot I_n \\
					& O_n & - I_n & \Ddots & \Vdots \\
					& \Vdots & \ddots & - I_n & 2 \cdot I_n \\
					& O_n & \Cdots & O_n & - I_n
					\CodeAfter \line{2-2}{3-3}
				\end{bNiceArray}
			\end{align}
		\end{minipage}
	\end{center}
\end{tcolorbox}

{\footnotesize
\begin{tcolorbox}
	[
		grow to left by = 2.65 cm,
		grow to right by = 2.65 cm,
		colback = white, 
		enhanced jigsaw, 
		sharp corners,
		boxrule = 0.01 pt,
		toprule = 0.01 pt,
		bottomrule = 0.01 pt
	]
	\begin{center}
		\begin{minipage}{0.5 \textwidth}
			\begin{align} \label{eq:Matrix M_E^c}
				\NiceMatrixOptions{ xdots/shorten = 0.85 em, code-for-first-row = \color{GreenLighter2}, code-for-first-col = \color{RedPurple} }
				H_{E^c} \mapsto M_{E^c} = \qquad
				\begin{bNiceArray}{cccc}[ first-row, first-col, rules/color = WordAquaLighter40, rules/width = 0.2 mm ]
					& \Ldots[ line-style = {solid, <->}, shorten = 0pt ]^{n^2 \text{ columns}} & \\
					\Vdots[ line-style = {solid, <->} ]_{n^2 \text{ rows}} 
					& W_{ \overline{ (1, 1) } } & W_{ \overline{ (1, 2) } } & \Cdots & W_{ \overline{ (1, n) } } \\
					& W_{ \overline{ (2, 1) } } & W_{ \overline{ (2, 2) } } & \Ddots & \Vdots \\
					& \Vdots & \Ddots & W_{ \overline{ (n-1, n-1) } } &  W_{ \overline{ (n-1, n) } } \\
					& W_{ \overline{ (n, 1) } } & \Cdots & W_{ \overline{ (n, n-1) } } & W_{ \overline{ (n, n) } }
					\CodeAfter \line{2-2}{3-3}
				\end{bNiceArray}
			\end{align}
		\end{minipage}
		\begin{minipage}{0.495 \textwidth}
			\begin{align} \label{eq:Matrix M_W}
				\NiceMatrixOptions{ xdots/shorten = 0.85 em, code-for-first-row = \color{GreenLighter2}, code-for-first-col = \color{RedPurple} }
				H_{W} \mapsto M_{W} = \qquad
				\begin{bNiceArray}{cccc}[ first-row, first-col, rules/color = WordAquaLighter40, rules/width = 0.2 mm ]
					& \Ldots[ line-style = {solid, <->}, shorten = 0pt ]^{n^2 \text{ columns}} & \\
					\Vdots[ line-style = {solid, <->} ]_{n^2 \text{ rows}} 
					& W_{ ( 1, 1 ) } & W_{ ( 1, 2 ) } & \Cdots & W_{ ( 1, n ) } \\
					& W_{ ( 2, 1 ) } & W_{ ( 2, 2 ) } & \Ddots & \Vdots \\
					& \Vdots & \Ddots & W_{ ( n - 1, n - 1 ) } &  W_{ ( n - 1, n ) } \\
					& W_{ ( n, 1 ) } & \Cdots & W_{ ( n, n - 1 ) } & W_{ ( n, n ) }
					\CodeAfter \line{2-2}{3-3}
				\end{bNiceArray}
			\end{align}
		\end{minipage}
	\end{center}
\end{tcolorbox}
}

In view of the relations (\ref{eq:Matrix M_v_p})--(\ref{eq:Matrix M_W}), the sum of the $3$ Hamiltonians $H_{v, p} + H_{p, v} + H_{E^c}$ can be associated with the linear expression $2 n + X^{\top} \left( M_{v, p} + M_{p, v} + M_{E^c} \right) X$, where $X$ is the column vector of the binary variables given by (\ref{eq:n^2-Dimensional Vector X}). The constant term $2 n$ is irrelevant in the whole quantum annealing process, and can be omitted. By defining for succinctness the matrix $M_{HCP}$, the Hamiltonian $H_{HCP}$ can be expressed in matrix form as shown below.

\begin{center}
	\begin{minipage}{0.45 \textwidth}
		\begin{align}
			H_{HCP} = X^{\top} M_{HCP} X \label{eq:HCP QUBO Matrix Form}
		\end{align}
	\end{minipage}
	\begin{minipage}{0.45 \textwidth}
		\begin{align}
			M_{HCP} = c_1 \left( M_{v, p} + M_{p, v} + M_{E^c} \right) \label{eq:Matrix M_HCP}
		\end{align}
	\end{minipage}
\end{center}

Analogously, $H_{TSP}$ can be cast in a compact matrix form with the help of the auxiliary matrix $M_{TSP}$.

\begin{center}
	\begin{minipage}{0.45 \textwidth}
		\begin{align}
			H_{HCP} = X^{\top} M_{TSP} X \label{eq:TSP QUBO Matrix Form}
		\end{align}
	\end{minipage}
	\begin{minipage}{0.45 \textwidth}
		\begin{align}
			M_{TSP} = M_{HCP} + c_2 M_{W} \label{eq:Matrix M_TSP}
		\end{align}
	\end{minipage}
\end{center}

\begin{example} \label{xmp:Matrix QUBO Example}
	To demonstrate how the previous machinery can be utilized, we return to the graph $G_{1}$ of Example \ref{xmp:Standard QUBO Example} and express the HCP and TSP Hamiltonians as matrices. For clarity of presentation and simplicity we will take the $2$ positive constants $c_1$ and $c_2$ equal to $1$. To emphasize the fact the $X$ and the matrices computed in this example refer to the particular graph $G_{1}$, we use the notation $X^{ G_{1} }$, $M_{v, p}^{ G_{1} }$, $M_{p, v}^{ G_{1} }$, $M_{E^c}^{ G_{1} }$, $M_{W}^{ G_{1} }$, $M_{HCP}^{ G_{1} }$ and $M_{TSP}^{ G_{1} }$ when referring to them. In this example, the column vector $X^{ G_{1} }$, containing the $n^2 = 16$ binary variables, is listed below. Note that in order to save space we actually show $\left( X^{ G_{1} } \right)^{\top}$.

	{\small
		\begin{align} \label{eq:Example Vector X}
			\setcounter{MaxMatrixCols}{17}
			\left( X^{ G_{1} } \right)^{\top}
			&=
			\left[
			\begin{NiceArray}{rrrrrrrr}[ margin = 1 mm ]
				x_{1, 1} & x_{1, 2} & x_{1, 3} & x_{1, 4} & x_{2, 1} & x_{2, 2} & x_{2, 3} & x_{2, 4}
			\end{NiceArray}
			\right.
			\nonumber \\
			&\hspace{4.05 cm}
			\left.
			\begin{NiceArray}{rrrrrrrr}[ margin = 1 mm ]
				x_{3, 1} & x_{3, 2} & x_{3, 3} & x_{3, 4} & x_{4, 1} & x_{4, 2} & x_{4, 3} & x_{4, 4}
			\end{NiceArray}
			\right]
		\end{align}
	}

	Matrices $M_{v, p}^{ G_{1} }, M_{p, v}^{ G_{1} }, M_{E^c}^{ G_{1} }$ and $M_{W}^{ G_{1} }$, adopted to the specific graph $G_{1}$, now become as follows.

	{\footnotesize
		\begin{align} \label{eq:Example Matrix M_v_p}
			\setcounter{MaxMatrixCols}{17}
			M_{v, p}^{ G_{1} }
			= \
			\begin{bNiceArray}{rrrr|rrrr|rrrr|rrrr}[ left-margin = 1.4 mm, right-margin = 1.4 mm, rules/color = WordAquaLighter40, rules/width = 0.25 mm ]
				-1 & 2 & 2 & 2 & 0 & 0 & 0 & 0 & 0 & 0 & 0 & 0 & 0 & 0 & 0 & 0 \\
				0 & -1 & 2 & 2 & 0 & 0 & 0 & 0 & 0 & 0 & 0 & 0 & 0 & 0 & 0 & 0 \\
				0 & 0 & -1 & 2 & 0 & 0 & 0 & 0 & 0 & 0 & 0 & 0 & 0 & 0 & 0 & 0 \\
				0 & 0 & 0 & -1 & 0 & 0 & 0 & 0 & 0 & 0 & 0 & 0 & 0 & 0 & 0 & 0 \\
				\hline
				0 & 0 & 0 & 0 & -1 & 2 & 2 & 2 & 0 & 0 & 0 & 0 & 0 & 0 & 0 & 0 \\
				0 & 0 & 0 & 0 & 0 & -1 & 2 & 2 & 0 & 0 & 0 & 0 & 0 & 0 & 0 & 0 \\
				0 & 0 & 0 & 0 & 0 & 0 & -1 & 2 & 0 & 0 & 0 & 0 & 0 & 0 & 0 & 0 \\
				0 & 0 & 0 & 0 & 0 & 0 & 0 & -1 & 0 & 0 & 0 & 0 & 0 & 0 & 0 & 0 \\
				\hline
				0 & 0 & 0 & 0 & 0 & 0 & 0 & 0 & -1 & 2 & 2 & 2 & 0 & 0 & 0 & 0 \\
				0 & 0 & 0 & 0 & 0 & 0 & 0 & 0 & 0 & -1 & 2 & 2 & 0 & 0 & 0 & 0 \\
				0 & 0 & 0 & 0 & 0 & 0 & 0 & 0 & 0 & 0 & -1 & 2 & 0 & 0 & 0 & 0 \\
				0 & 0 & 0 & 0 & 0 & 0 & 0 & 0 & 0 & 0 & 0 & -1 & 0 & 0 & 0 & 0 \\
				\hline
				0 & 0 & 0 & 0 & 0 & 0 & 0 & 0 & 0 & 0 & 0 & 0 & -1 & 2 & 2 & 2 \\
				0 & 0 & 0 & 0 & 0 & 0 & 0 & 0 & 0 & 0 & 0 & 0 & 0 & -1 & 2 & 2 \\
				0 & 0 & 0 & 0 & 0 & 0 & 0 & 0 & 0 & 0 & 0 & 0 & 0 & 0 & -1 & 2 \\
				0 & 0 & 0 & 0 & 0 & 0 & 0 & 0 & 0 & 0 & 0 & 0 & 0 & 0 & 0 & -1
			\end{bNiceArray}
		\end{align}
		\begin{align} \label{eq:Example Matrix M_p_v}
			\setcounter{MaxMatrixCols}{17}
			M_{p, v}^{ G_{1} }
			= \
			\begin{bNiceArray}{rrrr|rrrr|rrrr|rrrr}[ left-margin = 1.4 mm, right-margin = 1.4 mm, rules/color = WordAquaLighter40, rules/width = 0.25 mm ]
				-1 & 0 & 0 & 0 & 2 & 0 & 0 & 0 & 2 & 0 & 0 & 0 & 2 & 0 & 0 & 0 \\
				0 & -1 & 0 & 0 & 0 & 2 & 0 & 0 & 0 & 2 & 0 & 0 & 0 & 2 & 0 & 0 \\
				0 & 0 & -1 & 0 & 0 & 0 & 2 & 0 & 0 & 0 & 2 & 0 & 0 & 0 & 2 & 0 \\
				0 & 0 & 0 & -1 & 0 & 0 & 0 & 2 & 0 & 0 & 0 & 2 & 0 & 0 & 0 & 2 \\
				\hline
				0 & 0 & 0 & 0 & -1 & 0 & 0 & 0 & 2 & 0 & 0 & 0 & 2 & 0 & 0 & 0 \\
				0 & 0 & 0 & 0 & 0 & -1 & 0 & 0 & 0 & 2 & 0 & 0 & 0 & 2 & 0 & 0 \\
				0 & 0 & 0 & 0 & 0 & 0 & -1 & 0 & 0 & 0 & 2 & 0 & 0 & 0 & 2 & 0 \\
				0 & 0 & 0 & 0 & 0 & 0 & 0 & -1 & 0 & 0 & 0 & 2 & 0 & 0 & 0 & 2 \\
				\hline
				0 & 0 & 0 & 0 & 0 & 0 & 0 & 0 & -1 & 0 & 0 & 0 & 2 & 0 & 0 & 0 \\
				0 & 0 & 0 & 0 & 0 & 0 & 0 & 0 & 0 & -1 & 0 & 0 & 0 & 2 & 0 & 0 \\
				0 & 0 & 0 & 0 & 0 & 0 & 0 & 0 & 0 & 0 & -1 & 0 & 0 & 0 & 2 & 0 \\
				0 & 0 & 0 & 0 & 0 & 0 & 0 & 0 & 0 & 0 & 0 & -1 & 0 & 0 & 0 & 2 \\
				\hline
				0 & 0 & 0 & 0 & 0 & 0 & 0 & 0 & 0 & 0 & 0 & 0 & -1 & 0 & 0 & 0 \\
				0 & 0 & 0 & 0 & 0 & 0 & 0 & 0 & 0 & 0 & 0 & 0 & 0 & -1 & 0 & 0 \\
				0 & 0 & 0 & 0 & 0 & 0 & 0 & 0 & 0 & 0 & 0 & 0 & 0 & 0 & -1 & 0 \\
				0 & 0 & 0 & 0 & 0 & 0 & 0 & 0 & 0 & 0 & 0 & 0 & 0 & 0 & 0 & -1
			\end{bNiceArray}
		\end{align}
	}

	Before we proceed, let us remark that a straightforward computation of the expressions $X^{\top} M_{v, p}^{ G_{1} } X$ and $X^{\top} M_{p, v}^{ G_{1} } X$ will produce exactly the same constraints (minus the constant terms) as those given in equations (\ref{eq:H_v_p Constraints}) and (\ref{eq:H_p_v Constraints}), respectively. $M_{v, p}^{ G_{1} }$ and $M_{p, v}^{ G_{1} }$ are rather generic, in the sense that they do not convey any specific information about the graph at hand. The topology of the graph is captured by $M_{E^c}^{ G_{1} }$ and $M_{W}^{ G_{1} }$. Since $G_{1}$ is complete, matrix $M_{E^c}^{ G_{1} }$ assumes a particularly simple form. As expected, by explicitly computing $X^{\top} M_{E^c}^{ G_{1} } X$ we obtain exactly the same $16$ constraints encountered in (\ref{eq:H_E^c Constraints for Complete Graph}).

	{\footnotesize
		\begin{align} \label{eq:Example Matrix M_E^c}
			\setcounter{MaxMatrixCols}{17}
			M_{E^c}^{ G_{1} }
			= \
			\begin{bNiceArray}{cccc|cccc|cccc|cccc}[ left-margin = 1.4 mm, right-margin = 1.4 mm, rules/color = WordAquaLighter40, rules/width = 0.25 mm ]
				0 & 1 & 0 & 1 & 0 & 0 & 0 & 0 & 0 & 0 & 0 & 0 & 0 & 0 & 0 & 0 \\
				0 & 0 & 1 & 0 & 0 & 0 & 0 & 0 & 0 & 0 & 0 & 0 & 0 & 0 & 0 & 0 \\
				0 & 0 & 0 & 1 & 0 & 0 & 0 & 0 & 0 & 0 & 0 & 0 & 0 & 0 & 0 & 0 \\
				0 & 0 & 0 & 0 & 0 & 0 & 0 & 0 & 0 & 0 & 0 & 0 & 0 & 0 & 0 & 0 \\
				\hline
				0 & 0 & 0 & 0 & 0 & 1 & 0 & 1 & 0 & 0 & 0 & 0 & 0 & 0 & 0 & 0 \\
				0 & 0 & 0 & 0 & 0 & 0 & 1 & 0 & 0 & 0 & 0 & 0 & 0 & 0 & 0 & 0 \\
				0 & 0 & 0 & 0 & 0 & 0 & 0 & 1 & 0 & 0 & 0 & 0 & 0 & 0 & 0 & 0 \\
				0 & 0 & 0 & 0 & 0 & 0 & 0 & 0 & 0 & 0 & 0 & 0 & 0 & 0 & 0 & 0 \\
				\hline
				0 & 0 & 0 & 0 & 0 & 0 & 0 & 0 & 0 & 1 & 0 & 1 & 0 & 0 & 0 & 0 \\
				0 & 0 & 0 & 0 & 0 & 0 & 0 & 0 & 0 & 0 & 1 & 0 & 0 & 0 & 0 & 0 \\
				0 & 0 & 0 & 0 & 0 & 0 & 0 & 0 & 0 & 0 & 0 & 1 & 0 & 0 & 0 & 0 \\
				0 & 0 & 0 & 0 & 0 & 0 & 0 & 0 & 0 & 0 & 0 & 0 & 0 & 0 & 0 & 0 \\
				\hline
				0 & 0 & 0 & 0 & 0 & 0 & 0 & 0 & 0 & 0 & 0 & 0 & 0 & 1 & 0 & 1 \\
				0 & 0 & 0 & 0 & 0 & 0 & 0 & 0 & 0 & 0 & 0 & 0 & 0 & 0 & 1 & 0 \\
				0 & 0 & 0 & 0 & 0 & 0 & 0 & 0 & 0 & 0 & 0 & 0 & 0 & 0 & 0 & 1 \\
				0 & 0 & 0 & 0 & 0 & 0 & 0 & 0 & 0 & 0 & 0 & 0 & 0 & 0 & 0 & 0
			\end{bNiceArray}
		\end{align}
	}

	By combining the matrices $M_{v, p}^{ G_{1} }, M_{p, v}^{ G_{1} }$ and $M_{E^c}^{ G_{1} }$ according to equations (\ref{eq:HCP QUBO Matrix Form}) and (\ref{eq:Matrix M_HCP}), where, as explained before, constant $c_1 = 1$ for simplicity, we derive the matrix $M_{HCP}^{ G_{1} }$ and the matrix form of the Hamiltonian $H_{HCP}^{ G_{1} }$.

	{\footnotesize
		\begin{align} \label{eq:Example Matrix M_HCP}
			\setcounter{MaxMatrixCols}{17}
			M_{HCP}^{ G_{1} }
			= \
			\begin{bNiceArray}{rrrr|rrrr|rrrr|rrrr}[ left-margin = 1.4 mm, right-margin = 1.4 mm, rules/color = WordAquaLighter40, rules/width = 0.25 mm ]
				-2 & 3 & 2 & 3 & 2 & 0 & 0 & 0 & 2 & 0 & 0 & 0 & 2 & 0 & 0 & 0 \\
				0 & -2 & 3 & 2 & 0 & 2 & 0 & 0 & 0 & 2 & 0 & 0 & 0 & 2 & 0 & 0 \\
				0 & 0 & -2 & 3 & 0 & 0 & 2 & 0 & 0 & 0 & 2 & 0 & 0 & 0 & 2 & 0 \\
				0 & 0 & 0 & -2 & 0 & 0 & 0 & 2 & 0 & 0 & 0 & 2 & 0 & 0 & 0 & 2 \\
				\hline
				0 & 0 & 0 & 0 & -2 & 3 & 2 & 3 & 2 & 0 & 0 & 0 & 2 & 0 & 0 & 0 \\
				0 & 0 & 0 & 0 & 0 & -2 & 3 & 2 & 0 & 2 & 0 & 0 & 0 & 2 & 0 & 0 \\
				0 & 0 & 0 & 0 & 0 & 0 & -2 & 3 & 0 & 0 & 2 & 0 & 0 & 0 & 2 & 0 \\
				0 & 0 & 0 & 0 & 0 & 0 & 0 & -2 & 0 & 0 & 0 & 2 & 0 & 0 & 0 & 2 \\
				\hline
				0 & 0 & 0 & 0 & 0 & 0 & 0 & 0 & -2 & 3 & 2 & 3 & 2 & 0 & 0 & 0 \\
				0 & 0 & 0 & 0 & 0 & 0 & 0 & 0 & 0 & -2 & 3 & 2 & 0 & 2 & 0 & 0 \\
				0 & 0 & 0 & 0 & 0 & 0 & 0 & 0 & 0 & 0 & -2 & 3 & 0 & 0 & 2 & 0 \\
				0 & 0 & 0 & 0 & 0 & 0 & 0 & 0 & 0 & 0 & 0 & -2 & 0 & 0 & 0 & 2 \\
				\hline
				0 & 0 & 0 & 0 & 0 & 0 & 0 & 0 & 0 & 0 & 0 & 0 & -2 & 3 & 2 & 3 \\
				0 & 0 & 0 & 0 & 0 & 0 & 0 & 0 & 0 & 0 & 0 & 0 & 0 & -2 & 3 & 2 \\
				0 & 0 & 0 & 0 & 0 & 0 & 0 & 0 & 0 & 0 & 0 & 0 & 0 & 0 & -2 & 3 \\
				0 & 0 & 0 & 0 & 0 & 0 & 0 & 0 & 0 & 0 & 0 & 0 & 0 & 0 & 0 & -2
			\end{bNiceArray}
		\end{align}
	}
	{\footnotesize
		\begin{align} \label{eq:Example Matrix H_HCP}
			\setcounter{MaxMatrixCols}{17}
			H_{HCP}^{ G_{1} }
			= \
			X^{\top} M_{HCP}^{ G_{1} } X
		\end{align}
	}

	In this example, the quantitative information contained in the adjacency matrix (\ref{eq:G_1 Adjacency Matrix}), which gives the topology of the graph and the weight of each edge, is incorporated in the $M_{W}^{ G_{1} }$ matrix. It is easy to ascertain that $X^{\top} M_{W}^{ G_{1} } X$ creates exactly the same $24$ constraints as those given by equation (\ref{eq:H_W Constraints for Complete Graph}).

	{\footnotesize
		\begin{align} \label{eq:Example Matrix M_W}
			\setcounter{MaxMatrixCols}{17}
			M_{W}^{ G_{1} }
			= \
			\begin{bNiceArray}{rrrr|rrrr|rrrr|rrrr}[ left-margin = 1.4 mm, right-margin = 1.4 mm, rules/color = WordAquaLighter40, rules/width = 0.25 mm ]
				0 & 0 & 0 & 0 &  0 & 30 &  0 &  0 &  0 & 42 &  0 &  0 &  0 & 12 &  0 &  0 \\
				0 & 0 & 0 & 0 &  0 &  0 & 30 &  0 &  0 &  0 & 42 &  0 &  0 &  0 & 12 &  0 \\
				0 & 0 & 0 & 0 &  0 &  0 &  0 & 30 &  0 &  0 &  0 & 42 &  0 &  0 &  0 & 12 \\
				0 & 0 & 0 & 0 & 30 &  0 &  0 &  0 & 42 &  0 &  0 &  0 & 12 &  0 &  0 &  0 \\
				\hline
				 0 & 30 &  0 &  0 & 0 & 0 & 0 & 0 &  0 & 20 &  0 &  0 &  0 & 34 &  0 &  0 \\
				 0 &  0 & 30 &  0 & 0 & 0 & 0 & 0 &  0 &  0 & 20 &  0 &  0 &  0 & 34 &  0 \\
				 0 &  0 &  0 & 30 & 0 & 0 & 0 & 0 &  0 &  0 &  0 & 20 &  0 &  0 &  0 & 34 \\
				30 &  0 &  0 &  0 & 0 & 0 & 0 & 0 & 20 &  0 &  0 &  0 & 34 &  0 &  0 &  0 \\
				\hline
				 0 & 42 &  0 &  0 &  0 & 20 &  0 &  0 & 0 & 0 & 0 & 0 &  0 & 35 &  0 &  0 \\
				 0 &  0 & 42 &  0 &  0 &  0 & 20 &  0 & 0 & 0 & 0 & 0 &  0 &  0 & 35 &  0 \\
				 0 &  0 &  0 & 42 &  0 &  0 &  0 & 20 & 0 & 0 & 0 & 0 &  0 &  0 &  0 & 35 \\
				42 &  0 &  0 &  0 & 20 &  0 &  0 &  0 & 0 & 0 & 0 & 0 & 35 &  0 &  0 &  0 \\
				\hline
				 0 & 12 &  0 &  0 &  0 & 34 &  0 &  0 &  0 & 35 &  0 &  0 & 0 & 0 & 0 & 0 \\
				 0 &  0 & 12 &  0 &  0 &  0 & 34 &  0 &  0 &  0 & 35 &  0 & 0 & 0 & 0 & 0 \\
				 0 &  0 &  0 & 12 &  0 &  0 &  0 & 34 &  0 &  0 &  0 & 35 & 0 & 0 & 0 & 0 \\
				12 &  0 &  0 &  0 & 34 &  0 &  0 &  0 & 35 &  0 &  0 &  0 & 0 & 0 & 0 & 0
			\end{bNiceArray}
		\end{align}
	}

	Finally, by adding the matrices $M_{HCP}^{ G_{1} }$ and $M_{W}^{ G_{1} }$ as per equation (\ref{eq:Matrix M_TSP}), and taking $c_2 = 1$ for simplicity, we derive the matrix $M_{TSP}^{ G_{1} }$ and the matrix form of the Hamiltonian $H_{TSP}^{ G_{1} }$.

	{\footnotesize
		\begin{align} \label{eq:Example Matrix M_TSP}
			\setcounter{MaxMatrixCols}{17}
			M_{TSP}^{ G_{1} }
			= \
			\begin{bNiceArray}{rrrr|rrrr|rrrr|rrrr}[ left-margin = 1.4 mm, right-margin = 1.4 mm, rules/color = WordAquaLighter40, rules/width = 0.25 mm ]
				-2 & 3 & 2 & 3 &  2 & 30 &  0 &  0 &  2 & 42 &  0 &  0 &  2 & 12 &  0 &  0 \\
				0 & -2 & 3 & 2 &  0 &  2 & 30 &  0 &  0 &  2 & 42 &  0 &  0 &  2 & 12 &  0 \\
				0 & 0 & -2 & 3 &  0 &  0 &  2 & 30 &  0 &  0 &  2 & 42 &  0 &  0 &  2 & 12 \\
				0 & 0 & 0 & -2 & 30 &  0 &  0 &  2 & 42 &  0 &  0 &  2 & 12 &  0 &  0 &  2 \\
				\hline
				 0 & 30 &  0 &  0 & -2 & 3 & 2 & 3 &  2 & 20 &  0 &  0 &  2 & 34 &  0 &  0 \\
				 0 &  0 & 30 &  0 & 0 & -2 & 3 & 2 &  0 &  2 & 20 &  0 &  0 &  2 & 34 &  0 \\
				 0 &  0 &  0 & 30 & 0 & 0 & -2 & 3 &  0 &  0 &  2 & 20 &  0 &  0 &  2 & 34 \\
				30 &  0 &  0 &  0 & 0 & 0 & 0 & -2 & 20 &  0 &  0 &  2 & 34 &  0 &  0 &  2 \\
				\hline
				 0 & 42 &  0 &  0 &  0 & 20 &  0 &  0 & -2 & 3 & 2 & 3 &  2 & 35 &  0 &  0 \\
				 0 &  0 & 42 &  0 &  0 &  0 & 20 &  0 & 0 & -2 & 3 & 2 &  0 &  2 & 35 &  0 \\
				 0 &  0 &  0 & 42 &  0 &  0 &  0 & 20 & 0 & 0 & -2 & 3 &  0 &  0 &  2 & 35 \\
				42 &  0 &  0 &  0 & 20 &  0 &  0 &  0 & 0 & 0 & 0 & -2 & 35 &  0 &  0 &  2 \\
				\hline
				 0 & 12 &  0 &  0 &  0 & 34 &  0 &  0 &  0 & 35 &  0 &  0 & -2 & 3 & 2 & 3 \\
				 0 &  0 & 12 &  0 &  0 &  0 & 34 &  0 &  0 &  0 & 35 &  0 & 0 & -2 & 3 & 2 \\
				 0 &  0 &  0 & 12 &  0 &  0 &  0 & 34 &  0 &  0 &  0 & 35 & 0 & 0 & -2 & 3 \\
				12 &  0 &  0 &  0 & 34 &  0 &  0 &  0 & 35 &  0 &  0 &  0 & 0 & 0 & 0 & -2
			\end{bNiceArray}
		\end{align}
	}
	{\footnotesize
		\begin{align} \label{eq:Example Matrix H_TSP}
			\setcounter{MaxMatrixCols}{17}
			H_{TSP}^{ G_{1} }
			= \
			X^{\top} M_{TSP}^{ G_{1} } X
		\end{align}
	}
\hfill $\triangleleft$
\end{example}

\section{Normalizing graph weights} \label{sec:Normalizing Graph Weights}

During the extensive experimental tests of various instances of the TSP, a certain pattern consistently emerged. The aforementioned pattern demonstrates emphatically the importance of the numerical values of the weights used in the TSP Hamiltonian. In almost every occasion where the weights of the graph were above a certain threshold, e.g., $10$, the annealer preferred to break one or more of the imposed validity constraints rather than to strictly adhere to the topology of the graph. In hindsight, one could say that this is indeed the rational and expected behavior. \emph{If the weight of an edge is higher than the constraint penalty, the annealer, in order to reach a lower energy state, ignores the constraint, thus converging to solutions that violate the topology of the graph}. On the bright side, the predictability of this behavior, gives us the opportunity to remedy this problem. The solution that worked best in our experimental tests was the \emph{min-max normalization} \cite{Han2012} applied to the weights of the matrix $M_{W}$ \emph{before} adding the constraints. After the normalization process, the weight values are still proportional but lower than the constraint penalties.

Specifically, given the $n^{2} \times n^{2}$ matrix $M_{W} = [ m_{i, j} ]$, we denote by $m_{min}$ and $m_{max}$ the minimal and maximal elements of $M_{W}$, where without loss of generality we assume that $m_{min} \neq m_{max}$. All elements of $M_{W}$ are modified according to the formula (\ref{eq:Min-Max Normalization}) and the resulting normalized matrix is designated by $N_{W}$. Therefore, by defining appropriately the normalized matrix $N_{TSP}$, encompassing all the normalized constraints, we derive the \emph{normalized} matrix form of the Hamiltonian $H_{TSP}$, as shown below. From now on we shall always employ this normalized matrix, without further mention.

\begin{center}
	\begin{minipage}{0.45 \textwidth}
		\begin{align}
			N_{TSP} = M_{HCP} + N_{W} \label{eq:Matrix N_TSP}
		\end{align}
	\end{minipage}
	\begin{minipage}{0.45 \textwidth}
		\begin{align}
			H_{HCP} = X^{\top} N_{TSP} X \label{eq:Normalized TSP QUBO Matrix Form}
		\end{align}
	\end{minipage}
\end{center}

\begin{figure}[H]
	\begin{tcolorbox}
		[
			grow to left by = 1.25 cm,
			grow to right by = 1.25 cm,
			colback = gray!03,
			enhanced jigsaw, 
			sharp corners,
			toprule = 0.1 pt,
			bottomrule = 0.1 pt,
			leftrule = 0.1 pt,
			rightrule = 0.1 pt,
			sharp corners,
			center title,
			fonttitle = \bfseries
		]
		\begin{center}
			\begin{minipage}[b]{0.475 \textwidth}
				\begin{align} \label{eq:Min-Max Normalization}
					N_{W} [ i, j ]  = \frac{ M_{W} [ i, j ] - m_{min} }{ m_{max} - m_{min} }
				\end{align}
				\caption{The min-max normalization formula.}
				\label{fig:Min-Max Normalization}
			\end{minipage}
			\hfill
			\begin{minipage}[b]{0.495 \textwidth}
				\centering
				\begin{algorithm}[H]
					\caption{Min-max normalization.}
					\label{alg:Min-Max Normalization Algorithm}
					$min \leftarrow$ minimal element of $M_{W}$ \\
					$max \leftarrow$ maximal element of $M_{W}$ \\
					\For{$i \leftarrow 2$ \KwTo $n$}{
						\For{$j \leftarrow 2$ \KwTo $n$}{
							\texttt{$N_{W} [ i, j ] \gets ( M_{W} [ i, j ] - min ) / ( max - min )$}
						}
					}
				\end{algorithm}
				\caption{The min-max normalization algorithm.}
				\label{fig:Min-Max Normalization Algorithm}
			\end{minipage}
		\end{center}
	\end{tcolorbox}
\end{figure}

\begin{example} \label{xmp:Graph Weights Normalization Example}
	We explain the normalization process by examining how it can be applied to the matrix $M_{W}^{ G_{1} }$, as given by (\ref{eq:Example Matrix M_W}). Attempting to solve the TSP using $H_{TSP}^{ G_{1} }$ given by equation (\ref{eq:Example Matrix H_TSP}), which is based on the $M_{TSP}^{ G_{1} }$ matrix of equation (\ref{eq:Example Matrix M_TSP}), will typically yield a solution that violates some integrity constraints. We can counteract this by applying the normalization Algorithm \ref{alg:Min-Max Normalization Algorithm}. The final normalized QUBO model is shown below.
	\begin{tcolorbox}
		[
			grow to left by = 2.750 cm,
			grow to right by = 2.750 cm,
			colback = gray!03,
			enhanced jigsaw, 
			sharp corners,
			toprule = 0.1 pt,
			bottomrule = 0.1 pt,
			leftrule = 0.1 pt,
			rightrule = 0.1 pt,
			sharp corners,
			center title,
			fonttitle = \bfseries
		]
		{\footnotesize
		\begin{align} \label{eq:Normalized Example Matrix H_TSP}
			\setcounter{MaxMatrixCols}{17}
			N_{TSP}^{ G_{1} }
			= \
			\begin{bNiceArray}{rrrr|rrrr|rrrr|rrrr}[ rules/color = WordAquaLighter40, rules/width = 0.25 mm ] 
				-2.00 & 3.00 & 2.00 & 3.00 & 2.00 & 0.71 & 0.00 & 0.71 & 2.00 & 1.00 & 0.00 & 1.00 & 2.00 & 0.29 & 0.00 & 0.29 \\
				0.00 & -2.00 & 3.00 & 2.00 & 0.00 & 2.00 & 0.71 & 0.00 & 0.00 & 2.00 & 1.00 & 0.00 & 0.00 & 2.00 & 0.29 & 0.00 \\
				0.00 & 0.00 & -2.00 & 3.00 & 0.00 & 0.00 & 2.00 & 0.71 & 0.00 & 0.00 & 2.00 & 1.00 & 0.00 & 0.00 & 2.00 & 0.29 \\
				0.00 & 0.00 & 0.00 & -2.00 & 0.00 & 0.00 & 0.00 & 2.00 & 0.00 & 0.00 & 0.00 & 2.00 & 0.00 & 0.00 & 0.00 & 2.00 \\
				\hline
				0.00 & 0.71 & 0.00 & 0.71 & -2.00 & 3.00 & 2.00 & 3.00 & 2.00 & 0.48 & 0.00 & 0.48 & 2.00 & 0.81 & 0.00 & 0.81 \\
				0.00 & 0.00 & 0.71 & 0.00 & 0.00 & -2.00 & 3.00 & 2.00 & 0.00 & 2.00 & 0.48 & 0.00 & 0.00 & 2.00 & 0.81 & 0.00 \\
				0.00 & 0.00 & 0.00 & 0.71 & 0.00 & 0.00 & -2.00 & 3.00 & 0.00 & 0.00 & 2.00 & 0.48 & 0.00 & 0.00 & 2.00 & 0.81 \\
				0.00 & 0.00 & 0.00 & 0.00 & 0.00 & 0.00 & 0.00 & -2.00 & 0.00 & 0.00 & 0.00 & 2.00 & 0.00 & 0.00 & 0.00 & 2.00 \\
				\hline
				0.00 & 1.00 & 0.00 & 1.00 & 0.00 & 0.48 & 0.00 & 0.48 & -2.00 & 3.00 & 2.00 & 3.00 & 2.00 & 0.83 & 0.00 & 0.83 \\
				0.00 & 0.00 & 1.00 & 0.00 & 0.00 & 0.00 & 0.48 & 0.00 & 0.00 & -2.00 & 3.00 & 2.00 & 0.00 & 2.00 & 0.83 & 0.00 \\
				0.00 & 0.00 & 0.00 & 1.00 & 0.00 & 0.00 & 0.00 & 0.48 & 0.00 & 0.00 & -2.00 & 3.00 & 0.00 & 0.00 & 2.00 & 0.83 \\
				0.00 & 0.00 & 0.00 & 0.00 & 0.00 & 0.00 & 0.00 & 0.00 & 0.00 & 0.00 & 0.00 & -2.00 & 0.00 & 0.00 & 0.00 & 2.00 \\
				\hline
				0.00 & 0.29 & 0.00 & 0.29 & 0.00 & 0.81 & 0.00 & 0.81 & 0.00 & 0.83 & 0.00 & 0.83 & -2.00 & 3.00 & 2.00 & 3.00 \\
				0.00 & 0.00 & 0.29 & 0.00 & 0.00 & 0.00 & 0.81 & 0.00 & 0.00 & 0.00 & 0.83 & 0.00 & 0.00 & -2.00 & 3.00 & 2.00 \\
				0.00 & 0.00 & 0.00 & 0.29 & 0.00 & 0.00 & 0.00 & 0.81 & 0.00 & 0.00 & 0.00 & 0.83 & 0.00 & 0.00 & -2.00 & 3.00 \\
				0.00 & 0.00 & 0.00 & 0.00 & 0.00 & 0.00 & 0.00 & 0.00 & 0.00 & 0.00 & 0.00 & 0.00 & 0.00 & 0.00 & 0.00 & -2.00
			\end{bNiceArray}
		\end{align}
		}
	\end{tcolorbox}

	After the normalization, the constraints imposed on the QUBO model by the weights of the TSP graph retain their respective importance while assuming a range from $0$ (for the minimum weight) to $1$ (for the maximum weight). Thus, we can impose the rest of our constraints without conflicts, as the higher values in the previous QUBO model no longer overshadow them, forcing the annealer to ignore them. \hfill $\triangleleft$
\end{example}

\begin{figure}[H]
	\begin{tcolorbox}
		[
			grow to left by = 0.0 cm,
			grow to right by = 0.0 cm,
			colback = white, 
			enhanced jigsaw, 
			sharp corners,
			toprule = 1.0 pt,
			bottomrule = 1.0 pt,
			leftrule = 0.1 pt,
			rightrule = 0.1 pt,
			sharp corners,
			center title,
			fonttitle = \bfseries
		]
		\centering
		\begin{minipage}{1.00 \textwidth}
			\centering
			\includegraphics[scale = 1.00, trim = {0 0 0cm 0}, clip]{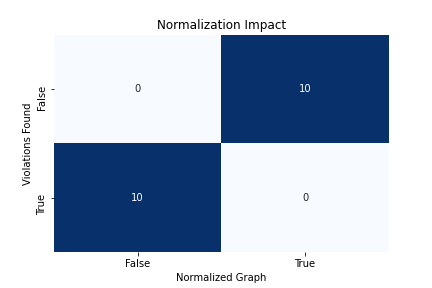}
			\caption{LeapHybridSampler returned a valid solution in all $10$ normalized instances, but failed to do so in all $10$ instances that weren't normalized.}
			\label{fig:Normalization Impact}
		\end{minipage}
	\end{tcolorbox}
\end{figure}

Figure \ref{fig:Normalization Impact} demonstrates the importance of the normalization process as a tool that amplifies the probability that the solutions produced by the annealer will not violate any of the constraints imposed by the QUBO model. To test its effectiveness We ran a total of $20$ experiments on D-Wave's LeapHybridSampler. The normalization procedure was applied to exactly half of the instances, e.g., $10$, and the other instances were executed without normalization. As can be seen in Figure \ref{fig:Normalization Impact}, LeapHybridSampler failed to yield a valid solution satisfying the graph topology in all $10$ instances that weren't normalized, whereas LeapHybridSampler succeeded in producing a valid solution in all $10$ normalized instances.

\section{Experimenting on D-Wave's QPU} \label{sec:Experimenting on D-Wave's QPU}

Our tests were conducted on two of D-Wave's available QPUs, on the {\tt Advantage\_System1.1} \cite{DWaveAdvantage2020}, which has been decommissioned since late 2021 \cite{Advantage1.1Decommissioned2021}), and on its newer version {\tt Advantage\_System4.1} \cite{DWaveAdvantage4.1.2021}.

\subsection{HCP using QPU} \label{subsec:HCP using QPU}

As explained in section \ref{sec:Introduction}, the Hamiltonian Cycle Problem (HCP) is the basis for the Traveling Salesman Problem (TSP), and its solution is an integral part of the solution of the TSP. We solved the HCP using both QPU systems available by D-Wave, namely {\tt Advantage\_system1.1} and {\tt Advantage\_system4.1}. The Flinders Hamiltonian Cycle Project (FHCP) provides many benchmarking graph datasets for both the HCP and the TSP \cite{FHCP2022}. We used their dataset of small graphs, namely the graph instances H\_6, H\_8, H\_10, H\_12, H\_14, generated by GENREG \cite{Meringer1999}. Their small number of nodes made them fit to run on D-Wave's Quantum Annealer's available qubits, without requiring the use of a hybrid solver. The average metrics that were obtained during the experiments are summarized in Table \ref{tbl:HCP using QPU}. The ``Nodes'' column contains the number of nodes of the graphs, the ``Solver'' column shows the selected QPU that was used to solve the problem, and the ``Runs'' column gives the total number of times the experiment was executed. The ``QAT'' column shows the QPU Access Time, that is the average time it took the Quantum Processing Unit to solve the problem over all runs. The ``Qubits'' column contains the average number of qubits used in the embedding for each run. Finally, the ``Success Rate'' column depicts the percentage of runs that provided a solution that didn't violate any of the imposed constraints.

\begin{table}[H]
	\renewcommand{\arraystretch}{1.5}
	\begin{tcolorbox}
		[
			grow to left by = 0.0 cm,
			grow to right by = 0.0 cm,
			colback = gray!03,
			enhanced jigsaw, 
			sharp corners,
			boxrule = 0.1 pt,
			toprule = 0.1 pt,
			bottomrule = 0.1 pt
		]
		\caption{This Table contains the average metrics that were obtained from solving HCP instances using D-Wave's QPU. \\}
		\label{tbl:HCP using QPU}
		\centering
		{\small
		\begin{tabular}
			{
				>{\centering\arraybackslash} m{1.25 cm} !{\vrule width 0.5 pt} >{\centering\arraybackslash} m{3.50 cm} !{\vrule width 0.5 pt} >{\centering\arraybackslash} m{1.25 cm} !{\vrule width 0.5 pt}
				>{\centering\arraybackslash} m{1.75 cm} !{\vrule width 0.5 pt}
				>{\centering\arraybackslash} m{1.75 cm} !{\vrule width 0.5 pt}
				>{\centering\arraybackslash} m{2.00 cm}
			}
			\Xhline{3\arrayrulewidth}
			\textbf{Nodes}
			&
			\textbf{Solver}
			&
			\textbf{Runs}
			&
			\textbf{QAT ($\pmb \mu$s)}
			&
			\textbf{Qubits}
			&
			\textbf{Success Rate (\%)}
			\\
			\Xhline{3\arrayrulewidth}
			$\phantom{0}6$ & {\tt Advantage\_system1.1} & $20$ & $126643.89$ & $126.83$ & $100.0$
			\\
			\Xhline{\arrayrulewidth}
			$\phantom{0}6$ & {\tt Advantage\_system4.1} & $20$ & $140461.42$ & $120.27$ & $100.0$
			\\
			\Xhline{\arrayrulewidth}
			$\phantom{0}8$ & {\tt Advantage\_system1.1} & $20$ & $147772.21$ & $388.25$ & $\phantom{0}15.0$
			\\
			\Xhline{\arrayrulewidth}
			$\phantom{0}8$ & {\tt Advantage\_system4.1} & $20$ & $160148.34$ & $372.60$ & $\phantom{0}35.0$
			\\
			\Xhline{\arrayrulewidth}
			$10$ & {\tt Advantage\_system1.1} & $20$ & $168190.20$ & $1010.20$ & $\phantom{00}0.0$
			\\
			\Xhline{\arrayrulewidth}
			$10$ & {\tt Advantage\_system4.1} & $20$ & $209123.62$ & $910.15$ & $\phantom{00}0.0$
			\\
			\Xhline{\arrayrulewidth}
			$12$ & {\tt Advantage\_system1.1} & $20$ & $170374.85$ & $2091.70$ & $\phantom{00}0.0$
			\\
			\Xhline{\arrayrulewidth}
			$12$ & {\tt Advantage\_system4.1} & $20$ & $243015.98$ & $1924.95$ & $\phantom{00}0.0$
			\\
			\Xhline{\arrayrulewidth}
			$14$ & {\tt Advantage\_system1.1} & $20$ & $188910.52$ & $3913.45$ & $\phantom{00}0.0$
			\\
			\Xhline{\arrayrulewidth}
			$14$ & {\tt Advantage\_system4.1} & $20$ & $272583.67$ & $3413.00$ & $\phantom{00}0.0$
			\\
			\Xhline{4\arrayrulewidth}
		\end{tabular}
		}
	\end{tcolorbox}
	\renewcommand{\arraystretch}{1.0}
\end{table}

As can be seen, violations of the imposed constraints start to appear when the number of nodes $n \geq 8$. This is in line with similar findings in a recent work \cite{Warren2021}. We also point out the clear improvements of {\tt Advantage\_system4.1} compared to {\tt Advantage\_system1.1}, both with respect to the success rate and the qubit usage. While {\tt Advantage\_system4.1} seems to be relatively slower, it more than makes up for it by offering better overall performance.

\begin{figure}[H]
	\begin{tcolorbox}
		[
			grow to left by = 0.0 cm,
			grow to right by = 0.0 cm,
			colback = white, 
			enhanced jigsaw, 
			sharp corners,
			toprule = 1.0 pt,
			bottomrule = 1.0 pt,
			leftrule = 0.1 pt,
			rightrule = 0.1 pt,
			sharp corners,
			center title,
			fonttitle = \bfseries
		]
		\centering
		\begin{minipage}{1.00 \textwidth}
			\centering
			\includegraphics[scale = 0.5, trim = {0 0 0cm 0}, clip]{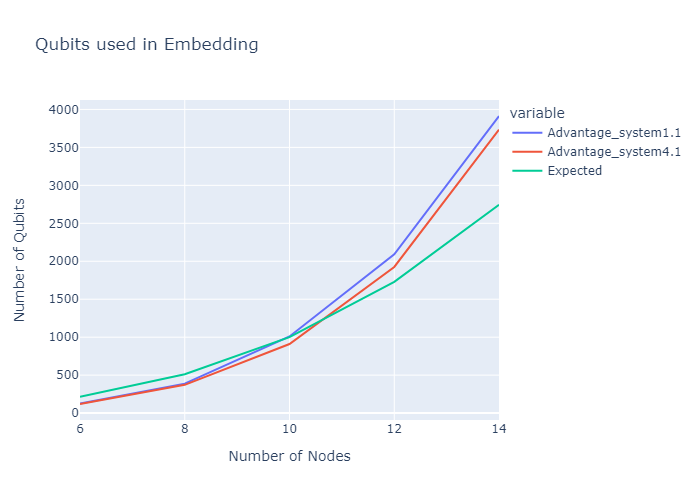}
			\caption{The number of qubits used in the embedding as a function of the number of nodes.}
			\label{fig:Qubits as a Function of Nodes}
		\end{minipage}
	\end{tcolorbox}
\end{figure}

Figure \ref{fig:Qubits as a Function of Nodes} gives the average number of qubits used in the embedding as a function of the number of nodes per graph. This visualization shows the graphs for the {\tt Advantage\_system1.1} solver, the {\tt Advantage\_system4.1}, and the expected qubits per instance. Both the {\tt Advantage\_system1.1} and the {\tt Advantage\_system4.1} demonstrate a rapidly increasing growth in the number of qubits as $n$ increases. One can also see that near and after the intersection points between the solvers' lines and the expected qubits' line is the point where the QPU starts breaking the imposed constraints. It is also noteworthy that the intersection point of {\tt Advantage\_system4.1} is to the right of {\tt Advantage\_system1.1}, something that reflects the first solver's superior success rate. In both cases, after the intersection point, every run on the solver returns a result that breaks at least one of the imposed constraints.

\begin{figure}[H]
	\begin{tcolorbox}
		[
			grow to left by = 0.0 cm,
			grow to right by = 0.0 cm,
			colback = white, 
			enhanced jigsaw, 
			sharp corners,
			toprule = 1.0 pt,
			bottomrule = 1.0 pt,
			leftrule = 0.1 pt,
			rightrule = 0.1 pt,
			sharp corners,
			center title,
			fonttitle = \bfseries
		]
		\centering
		\begin{minipage}{1.00 \textwidth}
			\centering
			\includegraphics[scale = 0.5, trim = {0 0 0cm 0}, clip]{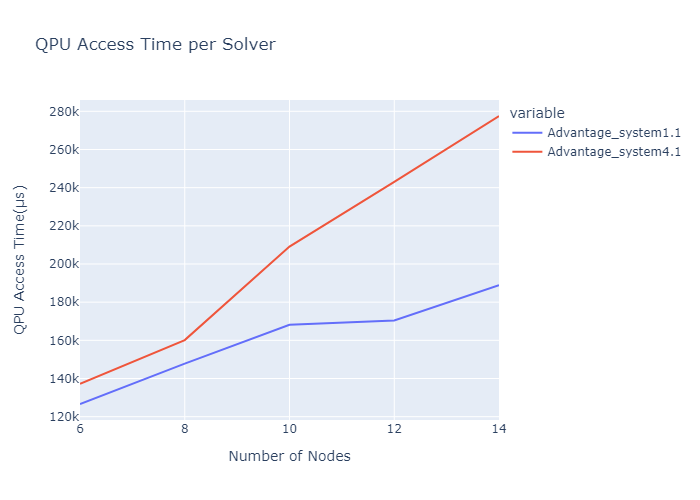}
			\caption{The average QPU Access Time as a function of the number of nodes.}
			\label{fig:QPU Access Time as a Function of Nodes}
		\end{minipage}
	\end{tcolorbox}
\end{figure}

Figure \ref{fig:QPU Access Time as a Function of Nodes} provides a comparison of the average QPU Access Time between the two solvers. It is evident that {\tt Advantage\_system4.1} needs more time to solve the problems, even if the difference is minuscule for real-world metrics. However, this relative loss of speed is compensated by the decrease of qubits in the embedding, meaning that {\tt Advantage\_system4.1} can handle bigger instances in a more satisfactory way.

\subsection{TSP using QPU} \label{subsec:TSP using QPU}

Moving from the HCP to the TSP is in principle quite straightforward, as their main difference is the introduction of weights on the graph. In our experiments, we relied on the TSPLIB. The TSPLIB is a library of TSP graph instances, commonly used for benchmarking algorithms that aim to solve the TSP \cite{Reinelt1991}. The TSPLIB instance with the least number of nodes is the {\tt burma14} graph, with just $n = 14$ nodes. Table \ref{tbl:TSP using QPU} below contains the average metrics that were gathered during the experiments. The ``Nodes'' column contains the number of nodes of the graphs, the ``Solver'' column shows the selected QPU that was used to solve the problem, and the ``Runs'' column gives the total number of times the experiment was executed. The ``QAT'' column shows the QPU Access Time, that is the average time it took the Quantum Processing Unit to solve the problem over all runs. The ``Qubits'' column contains the average number of qubits used in the embedding for each run. The ``Cost'' column shows the average cost of the solution path from each run. Finally, the ``Success Rate'' column depicts the percentage of runs that provided a solution that didn't violate any of the imposed constraints.

\begin{table}[H]
	\renewcommand{\arraystretch}{1.5}
	\begin{tcolorbox}
		[
			grow to left by = 0.75 cm,
			grow to right by = 0.75 cm,
			colback = gray!03,
			enhanced jigsaw, 
			sharp corners,
			boxrule = 0.1 pt,
			toprule = 0.1 pt,
			bottomrule = 0.1 pt
		]
		\caption{This Table contains the average metrics that were obtained from solving TSP instances using D-Wave's QPU. \\}
		\label{tbl:TSP using QPU}
		{\small
		\begin{tabular}
			{
				>{\centering\arraybackslash} m{1.25 cm} !{\vrule width 0.5 pt} >{\centering\arraybackslash} m{3.50 cm} !{\vrule width 0.5 pt} >{\centering\arraybackslash} m{1.25 cm} !{\vrule width 0.5 pt}
				>{\centering\arraybackslash} m{1.75 cm} !{\vrule width 0.5 pt}
				>{\centering\arraybackslash} m{1.75 cm} !{\vrule width 0.5 pt}
				>{\centering\arraybackslash} m{1.25 cm} !{\vrule width 0.5 pt}
				>{\centering\arraybackslash} m{2.00 cm}
			}
			\Xhline{3\arrayrulewidth}
			\textbf{Nodes}
			&
			\textbf{Solver}
			&
			\textbf{Runs}
			&
			\textbf{QAT ($\pmb \mu$s)}
			&
			\textbf{Qubits}
			&
			\textbf{Cost}
			&
			\textbf{Success Rate (\%)}
			\\
			\Xhline{3\arrayrulewidth}
			$14$ & {\tt Advantage\_system4.1} & $20$ & $282016.86$ & $4145.25$ & $4737.70$ & $0.0$
			\\
			\Xhline{4\arrayrulewidth}
		\end{tabular}
		}
	\end{tcolorbox}
	\renewcommand{\arraystretch}{1.0}
\end{table}

\begin{figure}[H]
	\begin{tcolorbox}
		[
			grow to left by = 0.0 cm,
			grow to right by = 0.0 cm,
			colback = white, 
			enhanced jigsaw, 
			sharp corners,
			toprule = 1.0 pt,
			bottomrule = 1.0 pt,
			leftrule = 0.1 pt,
			rightrule = 0.1 pt,
			sharp corners,
			center title,
			fonttitle = \bfseries
		]
		\centering
		\begin{minipage}{1.00 \textwidth}
			\centering
			\includegraphics[scale = 0.5, trim = {0 0 0cm 0}, clip]{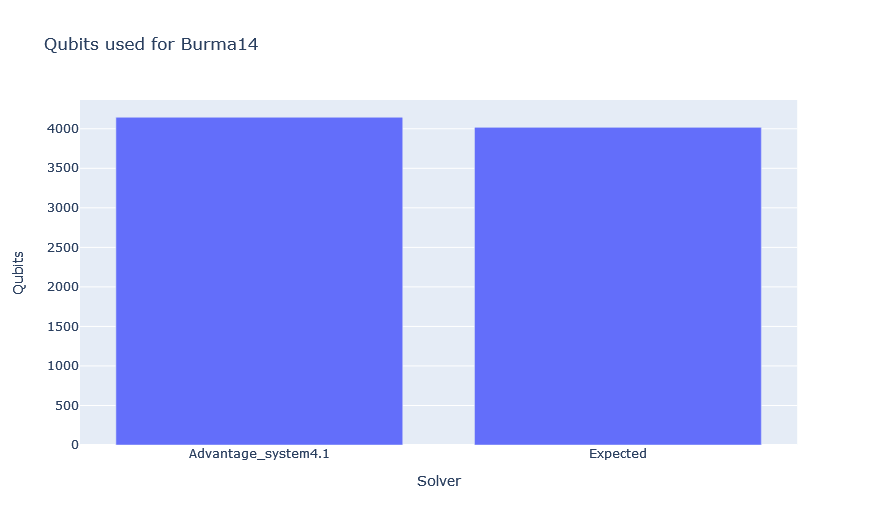}
			\caption{The average number of qubits used for {\tt Burma14} on the QPU.}
			\label{fig:Number of Qubits for Burma14}
		\end{minipage}
	\end{tcolorbox}
\end{figure}

The results in Table \ref{tbl:TSP using QPU} are quite similar to those obtained in the HCP case. Every run produced a solution that violated at least one of the integrity constraints. The qubit usage on the {\tt Advantage\_system4.1} is visually depicted in Figure \ref{fig:Number of Qubits for Burma14}. The number of qubits used is slightly higher than the optimal number given in Table \ref{tbl:Number of Constraints in Each Hamiltonian for a Complete Graph} of section \ref{sec:The Standard QUBO Formulation}. This, coupled with the improvement of the results produced by {\tt Advantage\_system4.1} shown in Table \ref{tbl:HCP using QPU}, is a clear indicator of the degree of technological progress achieved.

\subsection{The effect of graph connectivity on qubit usage} \label{subsec:The Effect of Graph Connectivity on Qubit Usage}

When comparing the results from the experiments, a consistent pattern emerged. Specifically, solving the HCP on graphs that were not complete, i.e., there were ``missing'' edges, always required more qubits than solving the HCP on complete graphs with the same number of nodes. We believe that this fact is the experimental validation of the theoretical results derived in Example \ref{xmp:Standard QUBO Example} and quantified in Tables \ref{tbl:Number of Constraints in Each Hamiltonian for a Complete Graph}, \ref{tbl:Number of Constraints in Each Hamiltonian for an Arbitrary Graph}, and \ref{tbl:Number of Constraints in Each Hamiltonian for a Simple Graph without Self-Loops}, which stipulate that when solving the HCP, incomplete graphs require more integrity constraints than complete graphs. In particular, the HCP on a graph with $n$ nodes and $k$ missing edges requires $n k$ additional quadratic constraints compared to a complete graph with $n$ nodes. To verify this experimentally, we created graphs of varying connectivity based on the FHCP $14$ nodes graph and measured the number of qubits required to solve the HCP with respect to the number of existing edges.

\begin{figure}[H]
	\begin{tcolorbox}
		[
			grow to left by = 0.0 cm,
			grow to right by = 0.0 cm,
			colback = white, 
			enhanced jigsaw, 
			sharp corners,
			toprule = 1.0 pt,
			bottomrule = 1.0 pt,
			leftrule = 0.1 pt,
			rightrule = 0.1 pt,
			sharp corners,
			center title,
			fonttitle = \bfseries
		]
		\centering
		\begin{minipage}{1.00 \textwidth}
			\centering
			\includegraphics[scale = 0.5, trim = {0 0 0cm 0}, clip]{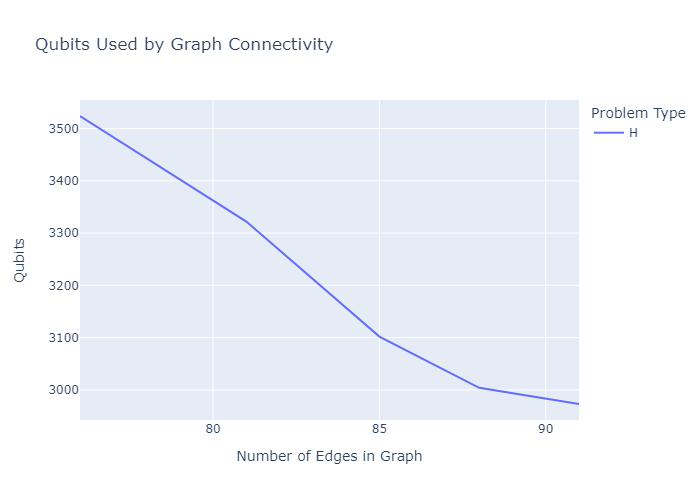}
			\caption{The average number of qubits used as a function of the number of edges in a graph with $14$ nodes.}
			\label{fig:Number of Qubits as a Function of the Number of Edges}
		\end{minipage}
	\end{tcolorbox}
\end{figure}

Figure \ref{fig:Number of Qubits as a Function of the Number of Edges} shows the number of qubits used in the embedding of the HCP Hamiltonian, as a function of the number of edges of the graph. It is evident that the more connected the graph is the fewer qubits its embedding needs. This is in full compliance with the theoretical analysis outlined in Table \ref{tbl:Number of Constraints in Each Hamiltonian for an Arbitrary Graph} of section \ref{sec:The Standard QUBO Formulation}.

\section{Experimenting on D-Wave's Leap’s Hybrid Solvers} \label{sec:Experimenting on D-Wave's Leap’s Hybrid Solvers}

To determine the viability of solving the HCP and TSP problems on a hybrid solver, we used D-Wave's LeapHybridSampler \cite{LHS2022} because it can sample arbitrarily big problems \cite{LHS2022BigProblems}.

\subsection{HCP using QPU-Hybrid} \label{subsec:HCP using QPU-Hybrid}

Running the HCP problems using D-Wave's LeapHybridSolver resulted in the metics found in Table \ref{tbl:HCP using LHS}. The ``Nodes'' column contains the number of nodes of the graphs, the ``Solver'' column shows the selected QPU that was used to solve the problem, and the ``Runs'' column gives the total number of times the experiment was executed. The ``QAT'' column shows the QPU Access Time, that is the average time it took the Quantum Processing Unit to solve the problem over all runs. The ``Run Time'' column contains is the average total time (in miliseconds) it took the solver to solve the problem instance. Finally, the ``Success Rate'' column depicts the percentage of runs that provided a solution that didn't violate any of the imposed constraints.

\begin{table}[H]
	\renewcommand{\arraystretch}{1.5}
	\begin{tcolorbox}
		[
			grow to left by = 1.25 cm,
			grow to right by = 1.25 cm,
			colback = gray!03,
			enhanced jigsaw, 
			sharp corners,
			boxrule = 0.1 pt,
			toprule = 0.1 pt,
			bottomrule = 0.1 pt
		]
		\caption{This Table contains the average metrics that were obtained from solving HCP instances using D-Wave's LeapHybridSampler. \\}
		\label{tbl:HCP using LHS}
		\centering
		{\small
		\begin{tabular}
			{
				>{\centering\arraybackslash} m{1.25 cm} !{\vrule width 0.5 pt} >{\centering\arraybackslash} m{7.00 cm} !{\vrule width 0.5 pt} >{\centering\arraybackslash} m{1.25 cm} !{\vrule width 0.5 pt}
				>{\centering\arraybackslash} m{1.75 cm} !{\vrule width 0.5 pt}
				>{\centering\arraybackslash} m{1.75 cm} !{\vrule width 0.5 pt}
				>{\centering\arraybackslash} m{2.00 cm}
			}
			\Xhline{3\arrayrulewidth}
			\textbf{Nodes}
			&
			\textbf{Solver}
			&
			\textbf{Runs}
			&
			\textbf{QAT ($\pmb \mu$s)}
			&
			\textbf{Run Time ($\pmb \mu$s)}
			&
			\textbf{Success Rate (\%)}
			\\
			\Xhline{3\arrayrulewidth}
			$\phantom{0}6$ & {\tt hybrid\_binary\_quadratic\_model\_version2} & $20$ & $66339.20$ & $2994650.00$ & $100.0$
			\\
			\Xhline{\arrayrulewidth}
			$\phantom{0}8$ & {\tt hybrid\_binary\_quadratic\_model\_version2} & $20$ & $ 64866.25$ & $2992684.90$ & $100.0$
			\\
			\Xhline{\arrayrulewidth}
			$10$ & {\tt hybrid\_binary\_quadratic\_model\_version2} & $20$ & $63423.32$ & $2994797.68$ & $100.0$
			\\
			\Xhline{\arrayrulewidth}
			$12$ & {\tt hybrid\_binary\_quadratic\_model\_version2} & $20$ & $49859.00$ & $3003757.63$ & $100.0$
			\\
			\Xhline{\arrayrulewidth}
			$14$ & {\tt hybrid\_binary\_quadratic\_model\_version2} & $20$ & $35612.10$ & $3008348.10$ & $100.0$
			\\
			\Xhline{4\arrayrulewidth}
		\end{tabular}
		}
	\end{tcolorbox}
	\renewcommand{\arraystretch}{1.0}
\end{table}

In stark contrast with the results obtained from running the problems directly on the QPU, as shown in Table \ref{tbl:HCP using QPU}, on the hybrid solver all the runs returned a valid solution that did not violate any of the imposed constraints. No matter the size of the problem, the hybrid solver always returns an acceptable solution.
Another noteworthy discrepancy is the variance of the values in the QPU Access Time column. It does not seem to follow any distribution or linear growth but varies wildly. The way D-Wave's hybrid solver work, it is possible that in some runs, the QPU Access Time will be $0$, as the computer might solve the problem without using the actual QPU \cite{LHS2022NoQPUTime}.

\subsection{TSP using QPU-Hybrid} \label{subsec:TSP using QPU-Hybrid}

Running the TSP problems using D-Wave's LeapHybridSolver resulted in the metrics found in Table \ref{tbl:TSP using LHS}.

The ``Nodes'' column contains the number of nodes of the graphs, the ``Solver'' column shows the selected QPU that was used to solve the problem, and the ``Runs'' column gives the total number of times the experiment was executed. The ``QAT'' column shows the QPU Access Time, that is the average time it took the Quantum Processing Unit to solve the problem over all runs. The ``Run Time'' column contains is the average total time (in miliseconds) it took the solver to solve the problem instance. The ``Cost'' column shows the average cost of the solution path from each run. Finally, the ``Success Rate'' column depicts the percentage of runs that provided a solution that didn't violate any of the imposed constraints. Table \ref{tbl:TSP using LHS} shows that while the QPU Access Time is an increasing function of $n$, the results are again different from table \ref{tbl:HCP using QPU} since the hybrid computer did not access the QPU every time. We particularly emphasize that the LeapHybridSolver successfully solved the {\tt gr120} problem from the TSPLIB, an instance with $120$ nodes. The fact that the resulting paths were far from optimal, can be attributed to the way LeapHybridSolver breaks the problems down to smaller instances, possibly losing important ``shortcut'' paths.

\begin{table}[H]
	\renewcommand{\arraystretch}{1.5}
	\begin{tcolorbox}
		[
		grow to left by = 1.50 cm,
		grow to right by = 1.50 cm,
		colback = gray!03,
		enhanced jigsaw, 
		sharp corners,
		boxrule = 0.1 pt,
		toprule = 0.1 pt,
		bottomrule = 0.1 pt
		]
		\caption{This Table contains the average metrics that were gathered from experiments on TSP instances using D-Wave's LeapHybridSampler. \\}
		\label{tbl:TSP using LHS}
		\centering
		{\small
		\begin{tabular}
			{
				>{\centering\arraybackslash} m{1.00 cm} !{\vrule width 0.5 pt} >{\centering\arraybackslash} m{6.25 cm} !{\vrule width 0.5 pt} >{\centering\arraybackslash} m{1.00 cm} !{\vrule width 0.5 pt}
				>{\centering\arraybackslash} m{1.50 cm} !{\vrule width 0.5 pt}
				>{\centering\arraybackslash} m{1.75 cm} !{\vrule width 0.5 pt}
				>{\centering\arraybackslash} m{1.25 cm} !{\vrule width 0.5 pt}
				>{\centering\arraybackslash} m{1.75 cm}
			}
			\Xhline{3\arrayrulewidth}
			\textbf{Nodes}
			&
			\textbf{Solver}
			&
			\textbf{Runs}
			&
			\textbf{QAT ($\pmb \mu$s)}
			&
			\textbf{Run Time ($\pmb \mu$s)}
			&
			\textbf{Cost}
			&
			\textbf{Success Rate (\%)}
			\\
			\Xhline{3\arrayrulewidth}
			$\phantom{0}17$ & {\tt hybrid\_binary\_quadratic\_model\_version2} & $20$ & $\phantom{0}14217.95$ & $\phantom{0}2995196.1$ & $\phantom{0}2417.15$ & $100.0$
			\\
			\Xhline{\arrayrulewidth}
			$\phantom{0}21$ & {\tt hybrid\_binary\_quadratic\_model\_version2} & $20$ & $\phantom{00}7181.00$ & $\phantom{0}3003955.9$ & $\phantom{0}3775.75$ & $100.0$
			\\
			\Xhline{\arrayrulewidth}
			$120$ & {\tt hybrid\_binary\_quadratic\_model\_version2} & $\phantom{0}5$ & $107764.40$ & $75028024.6$ & $25871.20$ & $100.0$
			\\
			\Xhline{4\arrayrulewidth}
		\end{tabular}
	}
	\end{tcolorbox}
	\renewcommand{\arraystretch}{1.0}
\end{table}

\begin{figure}[H]
	\begin{tcolorbox}
		[
		grow to left by = 0.0 cm,
		grow to right by = 0.0 cm,
		colback = white, 
		enhanced jigsaw, 
		sharp corners,
		toprule = 1.0 pt,
		bottomrule = 1.0 pt,
		leftrule = 0.1 pt,
		rightrule = 0.1 pt,
		sharp corners,
		center title,
		fonttitle = \bfseries
		]
		\centering
		\begin{minipage}{1.00 \textwidth}
			\centering
			\includegraphics[scale = 0.5, trim = {0 0 0cm 0}, clip]{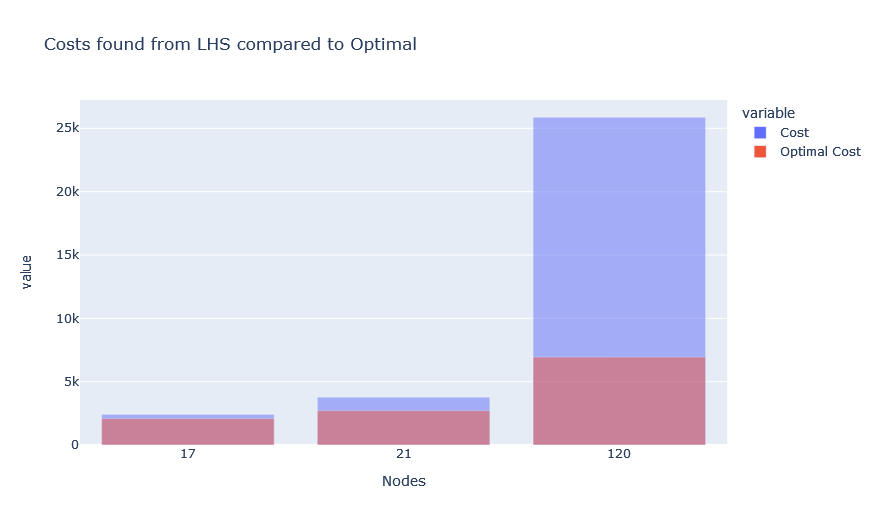}
			\caption{Comparison between the average costs of the solution produced by LeapHybridSampler and the optimal cost.}
			\label{fig:LHS TSP Cost vs. Optimal TSP Cost}
		\end{minipage}
	\end{tcolorbox}
\end{figure}

Figure \ref{fig:LHS TSP Cost vs. Optimal TSP Cost} presents the average cost of the solution produced by the LeapHybridSampler, compared to the optimal solution for each problem. One can see that as the number of nodes of the graph increases, the solution produced by LeapHybridSampler deviates from the optimal even more. Additionally, the average cost of the solutions reached by the annealer exhibits significant discrepancies compared to the optimal solutions. This is in line with the findings from a very recent study \cite{Warren2021} that also highlighted the difficulty the D-Wave solvers face in finding optimal solutions for TSP instances.

\section{Discussion and conclusions} \label{sec:Discussion and Conclusions}

In this work, we have considered various implementations, computational analyses, and comparisons on different versions of the D-Wave quantum annealer for the HCP and TSP problems. We extended our previous work by providing a complete matrix formulation for the HCP and TSP problems, and by introducing a normalization technique on the weight matrices. This technique helped avoid the constraint generated by the TSP graph's weights overshadowing the other imposed constraints, causing D-Wave's annealers to misbehave. We also ran some experiments on D-Wave's annealers and came to the following conclusions.

\begin{itemize}
	\item	D-Wave's {\tt Advantage\_system4.1} is more efficient than {\tt Advantage\_system1.1} in its use of qubits for solving a problem and provides more consistently correct solutions.
	\item	It is possible to run the {\tt Burma14} instance of the TSPLIB library on a quantum annealer using the aforementioned methods, although it can't provide a ``correct'' solution because of the annealer's limitations.
	\item	The more connected a graph is, the fewer qubits are needed for it to be solved by the quantum annealers, as fewer constraints need to be imposed.
	\item	Hybrid solvers always provide a correct solution to a problem and never break the constraints of a QUBO model, even for arbitrarily big problems.
\end{itemize}

The above conclusions are particularly significant as they indicate that quantum annealers can solve TSP and HCP problems. With the current advancements in quantum computing, the prospect of using a quantum annealer for solving real-world TSP problems looks very promising, especially if quantum computers continue to scale at the grade we see today.

Our computational results indicate that our method is very promising. We intend to continue this line of research and extend its usage to a variety of problem classes, with emphasis on other variations of TSP problems, such as the Traveling Salesperson Problem. Other possible improvements we intend to evaluate experimentally include the introduction of suitable ``phantom edges'' on graphs that are not fully connected, in order to circumvent the addition of more constraints, as well as experimenting with more normalization methods for the TSP graph weights.


\bibliographystyle{ieeetr}
\bibliography{ExperimentalAnalysisQuantumAnnealersHybridSolvers}

\end{document}